\documentstyle[12pt]{article}

\newcommand{\sect}[1]{\section{#1}\setcounter{equation}{0}}
\newcommand{\tr}{{\rm tr}\;}
\newcommand{\trr}{\tr R^2}
\newcommand{\trf}[1]{\tr F_{#1}^2 }
\newcommand{\npb}[3]{Nucl. Phys. {\bf B#1} (#2) #3}
\newcommand{\plb}[3]{Phys. Lett. {\bf #1B} (#2) #3}

\newcommand{\prl}[3]{Phys. Rev. Lett. {\bf #1} (#2) #3}
\newcommand{\mpl}[3]{Mod. Phys. Lett. {\bf A#1} (#2) #3}

\def\ZZ{{\bf Z}}
\def\Z{{\bf Z}}
\def\R{{\bf R}}
\def\T{{\bf T}}
\def\S{{\bf S}}
\begin{document}

\begin{titlepage}

\begin{flushright}
RU-96-16\\
NSF-ITP-96-21\\
CALT-68-2057\\
IASSNS-HEP-96/53\\
hep-th/9605184
\end{flushright}
\vskip 0.5in

\begin{center}

{\Large \bf Anomalies, Dualities, and Topology\\
of $D=6$ $N=1$ Superstring Vacua}
\vskip 1cm

{\bf Micha Berkooz,$^1$ Robert G. Leigh,$^1$ Joseph Polchinski,$^2$\\
John H. Schwarz,$^3$ Nathan Seiberg,$^1$} and {\bf Edward Witten$^4$}

\end{center}

\bigskip\noindent
{\it $^1$Department of Physics, Rutgers University, Piscataway, NJ\ \
08855-0849}\\
{\it $^2$Institute for Theoretical Physics, University of California,\\
\hspace*{3in} Santa Barbara,  CA 93106-4030}\\
{\it $^3$California Institute of Technology, Pasadena,
CA\ \ 91125}\\
{\it $^4$Institute for Advanced Study, Princeton, NJ 08540}

\bigskip\bigskip

\def\tilde{\widetilde}
\begin{abstract}
\baselineskip=16pt
\noindent
We consider various aspects of compactifications of the Type I/heterotic
$Spin(32)/\Z_2$ theory on K3.  One family of such compactifications
includes the standard embedding of the spin
connection in the gauge group, and is on the same moduli space as the
compactification of the heterotic $E_8\times E_8$ theory on K3 with
instanton numbers (8,16).  Another class, which includes an orbifold of
the Type I theory recently constructed by Gimon and Polchinski and whose
field theory limit involves some topological novelties,
is on the moduli space of the heterotic $E_8\times E_8$ theory on K3 with
instanton numbers (12,12).  These connections between $Spin(32)/\Z_2$
and  $E_8\times E_8$ models can be demonstrated
by T duality, and permit a better understanding of
non-perturbative gauge fields in the (12,12) model.
In the transformation
between $Spin(32)/\Z_2$ and $E_8\times E_8$ models, the
strong/weak coupling duality of the (12,12) $E_8\times E_8$
 model  is mapped to T duality in the Type I theory.  The gauge and
gravitational anomalies in the Type I theory are canceled by an
extension of the Green-Schwarz mechanism.

\end{abstract}\end{titlepage}
\newpage
\baselineskip=18pt

\sect{Introduction}

In the ongoing development of string duality, six-dimensional vacua with
$N=1$ supersymmetry have been a recent focus of interest.  In such
vacua, supersymmetry plus Lorentz invariance are not so strong as to
prevent interesting dynamics, but are strong enough to allow much of the
dynamics to be understood.  This was exploited, for example, in the
study of small instantons in the $SO(32)$ heterotic
string~\cite{sminst}.

There have been three broad categories of approaches developed for
obtaining six-dimensional string vacua with N=1 supersymmetry.  One
begins with the $E_8 \times E_8$ theory compactified on K3. Using the
recently developed 11-dimensional understanding of this string
theory~\cite{horava}, this approach has a natural extension to M-theory
compactified on ${\rm K3 }\times S^1 /\ZZ_2$.  The second approach begins
with the $SO(32)$ theory compactified on K3. Here, one has the option of
viewing the $SO(32)$ theory either as a heterotic string theory or a
Type I superstring theory, since they are nonperturbatively
equivalent~\cite{witusc}.
The third and most recent approach, which goes under the name of
F-theory, begins with the Type IIB superstring in ten
dimensions~\cite{vafaf}.  By associating a two-torus to the complex
scalar field of this theory, and allowing it to vary in non-trivial ways
over a four-dimensional base space B, one forms a Calabi-Yau space with
an elliptic fibration. Many different classes of vacua can be
constructed by each of these approaches, and in many cases it is clear
that constructions obtained by the various different approaches are dual
to one another -- {\it i.e.}, nonperturbatively equivalent.

Six-dimensional theories with N=1 supersymmetry contain four distinct
kinds of massless multiplets: the gravity multiplet, tensor multiplets,
vector multiplets, and hypermultiplets. The vector multiplets are
characterized by the choice of a gauge group and the hypermultiplets by
the choice of a representation of the group.  A tensor multiplet, which
is always a singlet of the group, contains a two-form potential with an
anti-self-dual field strength. Since the gravity multiplet contains a
two-form potential with a self-dual field strength, it is only possible
to give a manifestly covariant effective action when there is exactly
one tensor multiplet.  In this case the self-dual and anti-self-dual
tensors can be described together by a single two-form $B$ or a
three-form field strength $H$.  In traditional compactifications of
string theory from ten dimensions there is exactly one tensor multiplet.
Additional tensor multiplets are obtained in compactifications of
M-theory by adding small instantons, which can also be interpreted as
five-branes.  In the $SO(32)$ approach, on the other hand, the addition
of small instantons (or five-branes) leads to additional vector
multiplets and enhanced gauge symmetry, but no additional tensor
multiplets~\cite{sminst}.  In the F-theory constructions there are also
vacua without any tensor multiplet~\cite{seiwit}. Here we will only
consider vacua with one tensor multiplet.

Since all N=1 models in six dimensions are chiral, the cancellation of
anomalies is an important requirement. In particular, in the case of one
tensor multiplet, a necessary requirement is that the 8-form anomaly
polynomial factorize in the product of two 4-forms: $I_8 \sim X_4 \wedge
\tilde X_4$, where $X_4$ and $\tilde X_4$ have the structure
\begin{equation}
X_4 = {\rm tr} R^2 - \sum_{\alpha} v_{\alpha}  {\rm  tr} F_{\alpha}^2
\end{equation}
\begin{equation}
\tilde X_4 = {\rm tr} R^2 - \sum_{\alpha} \tilde v_{\alpha}  {\rm  tr}
F_{\alpha}^2
\end{equation}
and $\alpha$ labels the various factors in the gauge group. Actually, as
we will discuss in the next section, there can also be additional terms
in $I_8$ of the form $X_2 \wedge X_6$ when the group contains $U(1)$
factors. As is well-known, the four-forms also appear in the Bianchi
identity $dH = X_4$ and field equation $d*H = \tilde X_4$. It has been
realized recently that whenever any of the $\tilde v_{\alpha}$'s is
negative there is a value of the dilaton for which the coupling constant
of the corresponding gauge group diverges. This singularity is believed
to signal a phase transition associated with the appearance of
tensionless strings~\cite{seiwit, dlp}. This is a very interesting
phenomenon, which is the subject of much current discussion.  However,
in most of this paper we will focus on models (or parameter ranges)
for which this phenomenon does not occur.

In this paper we would like to elaborate on two recently-discussed
classes of $D=6$, $N=1$ vacua and to establish the connection between
them.  The first class is the $SO(32)$ Type~I string on $K3$ in the
$T^4/\Z_2$ orbifold limit~\cite{gimpol}.  The second is the $E_8\times
E_8$ heterotic string on K3, with symmetric embedding of the instantons
in the two $E_8$'s~\cite{dmw}. For convenience we will refer to these as
GP and DMW vacua respectively.  Along the way we will also consider the
$E_8\times E_8$ string with other embeddings, and other vacua of
the $Spin(32)/\Z_2$ heterotic string.

In ref. \cite{gimpol}, consistency conditions for open superstring
compactifications were studied, and the general solution was found for
the case of a $T^4/\Z_2$ orbifold.  Cancellation of tadpoles for
massless Ramond-Ramond 10-form and 6-form potentials required 32
Chan-Paton nine-brane indices and 32 five-brane indices, the five-branes
oriented so as to fill the 6-dimensional spacetime).\footnote {There is
a semantic problem in this subject.  In the GP models the five-brane
Chan-Paton index takes 32 values, but half can be regarded as the images
of the others under the orbifold $\Z_2$, and (as discussed in
refs.~\cite{sminst,gimpol}) half again can be regarded as images under
world-sheet parity $\Omega$.  Thus there are only 8 dynamical
five-branes, each having the minimum unit of 6-form charge.  So we will
refer to five-branes when, as will usually be the case, we mean the
dynamical objects, and to `indices' when we count the images
separately.} The gauge group associated with nine-branes was found to be
$U(16)$ (or a combination of unitary and symplectic subgroups, obtained
by adding Wilson lines) and the same for the five-branes with the
five-brane positions playing the role of the Wilson lines.

The world-sheet consistency conditions, from closure of the operator
product expansion and cancellation of one-loop divergences, were
conjectured to be a complete set.  If so, spacetime gauge
and gravitational anomaly cancellation, normally a tight constraint on the
spectra of $D=6$, $N=1$ Yang-Mills theories, will hold automatically.
In ref.~\cite{gimpol} the quartic terms in the spacetime anomalies were
found to cancel for all GP models.

In section~2 we study the full anomaly; as expected, the GP models are
(perturbatively) consistent, but the details are interesting. As is
familiar from refs.~\cite{gsmech}, in order that the anomalies be
canceled in the usual way (via tree-level exchange of the 2-form
$B_{\mu\nu}$), the anomaly 8-form for the non-abelian gauge fields
and gravity must factorize into a product of two
4-forms. In the consistent models of ref.~\cite{gimpol}, we find that
 part of the
anomaly polynomial involving {\it abelian} gauge fields
does {\it not} factorize in this fashion.  We identify an
extended version of the GS mechanism in which the anomalies involving
$U(1)$ gauge fields are
canceled by tree-level exchange of certain 0-form fields. These fields
are identified as R-R closed string twisted sector states. The requisite
inhomogeneous transformation of these fields under gauge transformations
leads to spontaneous symmetry breaking of some of the $U(1)$ factors
(just as occurs for the Green-Schwarz mechanism in four
dimensions~\cite{dsw}), so that some of the states identified as
massless in ref.~\cite{gimpol} receive masses of order the open string
coupling.

Although the world-sheet considerations of ref.~\cite{gimpol} evidently
guarantee a perturbatively consistent theory, we show that there is a
potential nonperturbative inconsistency.  For many of the GP models,
spinors of the gauge group cannot be defined.
Such spinors do not exist in
perturbation theory but do arise nonperturbatively as D-branes, so the
inconsistency appears first in this way.

In the GP construction there are two types of gauge groups: those
carried by open strings that connect pairs of five-branes and those
carried by open strings that connect pairs of nine-branes. As we will
explain, there is a T duality that interchanges the five-branes and
nine-branes and thereby interchanges the two types of open strings and
the gauge groups that they carry.

We will claim that these theories are on the same moduli space as the
class of vacua considered by DMW.  The latter exhibit an interesting
phenomenon: heterotic/heterotic string duality. That is, they have an
S-duality symmetry that interchanges perturbative heterotic strings with
non-perturbative heterotic strings.

We will argue that the GP and DMW are equivalent under a sort of
`U-duality' that maps the S-duality of the DMW description to the T
duality of the GP description.  This is an example of the widespread
phenomenon of ``duality of dualities."  Several pieces of evidence
support the proposed correspondence between these models.  One of these
is a comparison of the structure of the factorized anomaly polynomials
in each case. Another is the construction of solitonic strings in the GP
setup with the correct properties to be identified with the two kinds of
heterotic strings in the DMW description.  Finally, we will exhibit at
the end of this paper a direct proof of this relation using T duality
between certain orbifolds.

By integrating the Bianchi identity over K3 one learns that,
of $n_1$ and $n_2$ are the number of instantons in the two
factors, then  $E_8 \times E_8$ compactifications must satisfy
$n_1 +n_2 +n_5 =24$. As we have already said, setting $n_5$, the number
of $E_8 \times E_8$ fivebranes, to zero ensures that there is just one
tensor multiplet.  The DMW models correspond to the symmetric embedding
$n_1=n_2=12$. In the case of the $SO(32)$ theory compactified on K3,
there is a similar requirement, namely $n_1 +n_5 =24$, where $n_1$ is
the number of instantons embedded in the $SO(32)$ group and $n_5$ is the
number of $SO(32)$ five-branes (magnetic duals of the heterotic string
in ten dimensions).  Perturbative string vacua have $n_5 =0$, of course,
but by now we have become accustomed to considering non-perturbative
possibilities.  The GP construction uses the K3 orbifold $T^4/\ZZ_2$,
rather than a smooth K3. It turns out that the 16 orbifold singularities
each contain a ``hidden'' instanton, a fact that we will demonstrate by
studying the behavior upon
blowing up the singularity.  The  models constructed by GP
satisfy the counting rule by introducing eight five-branes in addition
to the hidden instantons.

In section~3, we use $D=10$ heterotic - Type I duality~\cite{witusc} to
relate the GP models to the $SO(32)$ heterotic string on K3.  From the
anomaly polynomial one learns that the nine-brane gauge group maps to a
perturbative gauge symmetry of the heterotic theory, and the five-brane
gauge group to a non-perturbative gauge symmetry.  It is therefore not
surprising that Type~I T-duality, which interchanges the nine- and
five-branes (for a review see ref.~\cite{pcj}), maps to the heterotic
weak/strong duality discussed in ref.~\cite{duff} and realized in the
DMW construction.  As discussed in ref.~\cite{duff}, the dual heterotic
string is a wrapped five-brane; in the Type~I theory we construct both
the heterotic string and its dual as D-branes~\cite{dbrane}.  Finally,
we show that for one particular GP model, the heterotic dual can be
constructed as a free orbifold.

In section~4 we consider the effect of blowing up the $\Z_2$ fixed
points by turning on twisted sector fields.  A count of instanton number
implies that there must be one instanton hidden at each fixed point.
The gauge bundle that results when the fixed point is blown up is a
$Spin(32)/\Z_2$ bundle but not an $SO(32)$ bundle.  That is, fields in
the vector representation of $SO(32)$ cannot be defined.  This is not an
inconsistency because the Type~I string has only tensor representations
(in perturbation theory) and one class of spinor representation
(nonperturbatively).\footnote{In the heterotic dual the same
representations appear, but all perturbatively.} A calculation of the
dimension of instanton moduli space agrees with the GP spectrum.

In section~5 we consider $T$-duality between the $SO(32)$ and $E_8\times
E_8$ heterotic strings on $K3$.  We first classify $\Z_2$ subgroups of
$Spin(32)/\Z_2$ and $E_8\times E_8$.  We then discuss $Spin(32)/\Z_2$
and $E_8\times E_8$ bundles (of instanton number 24) on K3 and the
possible $T$-dualities between them.  We resolve a puzzle remaining from
ref.~\cite{dmw}.  The nonperturbative gauge symmetries of the DMW model,
which had been tentatively attributed to small $E_8\times E_8$
instantons, in fact arise from small $Spin(32)/\Z_2$ instantons in the
$T$-dual theory.  Finally, we construct explicitly the $T$-duality
between the $SO(32)$ and $E_8\times E_8$ strings on $T^4/\Z_2$
orbifolds, with various embeddings of $\Z_2$ in the gauge groups, and in
particular complete the connection between the GP and DMW models.

Although we shall not explore the subject here, we should point out that
this class of models has also been constructed in
F-theory~\cite{vafamorrison}. The key there is to choose the base space of
the elliptic fibration to be ${\bf P_1} \times {\bf P_1}$. In this
description the inversion of heterotic string coupling
 is realized geometrically as the
interchange of the two ${\bf P_1}$ factors.

\sect{Anomalies}

\subsection{Anomalies in the $U(16)\times U(16)$ Model}

Reference \cite{gimpol} constructed a class of $N=1$ six-dimensional
Type~I superstring vacua with maximal gauge group $U(16)_9\times
U(16)_5$. The first $U(16)_9$ factor comes from nine-branes while the
second arises when all 8 five-branes are at a single fixed point of the
orbifold. The massless spectrum includes the hypermultiplets
\begin{eqnarray}
\label{spectr}
2\times &&\left( {\bf 120,1}\right)_{(2/4,0)}   \nonumber\\
1\times &&\left( {\bf 16,16}\right)_{(1/4,1/4)} \nonumber\\
2\times &&\left( {\bf 1, 120}\right)_{(0,2/4)}  \\
4\times &&\left( {\bf 1,1}\right)_{(0,0)}               \nonumber\\
16\times &&\left( {\bf 1,1}\right)_{(0,0)}              \nonumber
\end{eqnarray}
where the subscripts refer to the $U(1)$ charges.\footnote {The $U(1)$
charges are obtained by noting that the endpoints transform as a ${\bf
16} + \overline{\bf 16}$ of $U(16)$; the factor of 1/4 is included in
order to give a canonical normalization.} In this section we will
illustrate the cancellation of anomalies for this special case of maximal
gauge group, and in the next we will consider the general case.

In order for the anomalies of the theory (\ref{spectr}) to be canceled
by the standard GS mechanism \cite{gsmech}, it is necessary that the
anomaly 8-form factorize into a product of two 4-forms. The anomaly is
then canceled, in the low energy theory, by the appearance of a
gauge-variant counterterm coupling the field $B_{\mu\nu}$ to one of the
two 4-forms.

Recall that for N=1 models in six dimensions cancellation of the
coefficient of the tr~$R^4$ term in the anomaly polynomial requires that
$n_H - n_V = 273 - 29 n_T$, where $n_H$, $n_V$, and $n_T$ are the number
of hyper multiplets, vector multiplets, and tensor multiplets,
respectively.  In this paper we will only be considering models with
$n_T =1$. In addition, the hypermultiplet representations have to appear
in such a way as to ensure that all tr~$F^4$ terms can be eliminated.
In the theory with the field content given above, the $\tr\,\, R^4$ and
$\tr\,\, F^4$ terms do cancel, and the full gravitational
+ gauge anomaly may be written, using standard formulas,
 \cite{alvwit,gsw}, in the form
\begin{equation}
\label{anomaly}
I_8= X_4^{(9)}\wedge X_4^{(5)} + X_2\wedge X_6^{(9)}+ X'_2\wedge
X_6^{(5)}.
\end{equation}
The last two terms can appear only in the presence of $U(1)$ factors,
because they involve tr~$F$, which is nonvanishing for $U(1)$ only.  The
various forms appearing here are
\begin{eqnarray}
X_4^{(a)}=&&\trf{a}-{1\over 2}\trr      \nonumber\\
X_6^{(a)}=&&{8\over 3} \left( -{1\over 4}\trr\cdot\tr F_a
+\tr F_a^3-{9\over 4}\tr F_a\cdot\trf{a} +{3\over2}(\tr
F_a)^3\right)\nonumber\\
X_2=&& 4\;\tr F_9 + \tr F_5                     \\
X_2'=&& 4\;\tr F_5 + \tr F_9                    \nonumber
\end{eqnarray}
where $F_a$ with $a=5,9$ refer to the two $U(16)$ field strength
two-forms in the fundamental (16-dimensional) representation. A term of
the form tr~$F_a$ only involves the $U(1)$ subalgebra.  We note that
this is not of the usual factorized form, and thus the full anomaly
cannot be canceled by exchange of a two-form, which involves
GS interactions of the familiar form
\begin{equation}
\label{counterGS}
\Gamma_{c.t.}=\int B_2\wedge X_4^{(5)}+a_0\int X_3^{(9)}\wedge X_3^{(5)}
\end{equation}
for some (scheme-dependent) constant $a_0$.
The $B_2$ field is the 2-form with
gauge invariant field strength $H=dB_2-X_3^{(9)}$
where $X_4^{(a)}=dX_3^{(a)}$.

The counterterm (\ref{counterGS}) succeeds in canceling the first term
in (\ref{anomaly}). The remaining two terms may be canceled by
additional counterterms (up to gauge-dependent terms)
\begin{equation}
\label{counterBL}
\Gamma'_{c.t.}=
\int B_0^{(9)}\; X_6^{(9)}+\int B_0^{(5)}\; X_6^{(5)}
\end{equation}
if we assign the anomalous transformation laws:
\begin{eqnarray}
\label{anomshift}
B_0^{(9)}\rightarrow B_0^{(9)}+ 4\epsilon_9+\epsilon_5  \\
B_0^{(5)}\rightarrow B_0^{(5)}+ 4\epsilon_5+\epsilon_9  \nonumber
\end{eqnarray}
under the $U(1)$ gauge transformations $A_a\rightarrow A_a +
d\epsilon_a$.  It follows that the scalar fields $B_0$ must appear in
the low energy Lagrangian in the combinations $dB_0^{(a)}-q_{a,b}
A_{(b)}$, where $q_{a,b}$ are the coefficients of the shifts in
(\ref{anomshift}).  The inhomogeneity of the transformation
laws~(\ref{anomshift}) implies that the two $U(1)$'s are spontaneously
broken, just as in four-dimensional theories with anomalous
$U(1)$'s where a similar mechanism appears~\cite{dsw}. The
unbroken symmetry is therefore only
$SU(16)\times SU(16)$. The $B_0$ fields are linear combinations of R-R
twisted-sector scalars, which we will refer to as $\phi_I$, $I=1,\ldots,
16$ labeling the orbifold fixed points.  By examining string
amplitudes, we may identify the fields $B_0$ more concretely and deduce
their kinetic terms and anomalous couplings (\ref{counterBL}). These all
appear at disk order, as does the standard GS term (\ref{counterGS}).
In the next section, we will see that these couplings arise also in the
(equivalent) boundary-state tadpole formalism.
It is easy to infer from diagrams that the $B_0$ fields that cancel the
$U(1)$ anomalies are twisted sector R-R states.  Consider the mixing of
a scalar mode with a ${U(1)}$ gauge boson.  The Chan-Paton factor of the
$U(1)$ gauge boson is a matrix in the algebra of $SO(32)$ of the form:
\begin{equation}
\label{chpat}
\lambda\equiv M=\left(\matrix{0&I_{16}\cr -I_{16}&0}\right)
\end{equation}
($I_{16}$ is a $16\times 16$ unit matrix).
The orbifold projection $R$ acts on the Chan-Paton indices as matrices
$\gamma_{R,a}$, which are identical in form to $M$; again the index $a$
denotes either the nine- or five-brane sector. Consider first a coupling
of an untwisted closed-string mode to a $U(1)$ gauge boson. Such an
amplitude is proportional to $\tr \lambda$, and hence it vanishes
trivially.  However, insertion of a twist operator in the interior of
the disk creates a cut which must be taken to run from its insertion
point to the boundary.  The fields jump across the cut by the orbifold
operation $R$, which includes the matrix $\gamma_{R,a}$ on the
Chan-Paton degrees of freedom.  The disk amplitude is then proportional
to $\tr\gamma_{R,a}\lambda \neq 0$. The R-R twisted state associated to
a given fixed point then couples to the $U(1)$ at that fixed point,
while each of the sixteen R-R twisted states couple to the $U(1)_9$
gauge boson. We can therefore identify
\begin{eqnarray}
\label{RRfields}
B_0^{(5)} & = & \phi_{1}\\
B_0^{(9)} & = & {1\over 4}\sum_{I=1}^{16} \phi_I.\nonumber
\end{eqnarray}
This ensures that eqs. (\ref{anomshift}) are consistent, if
$\phi_1$ refers to the special fixed point and each of the other 15
twisted fields shift by $\phi_I\rightarrow\phi_I+\epsilon_9$. These
rules generalize in a straightforward way to all other models in Ref.
\cite{gimpol}, as we will see in the next section.  Incidentally, the
4-to-1 ratio of charges in (\ref{anomshift}) has a simple explanation:
given the natural normalization of fields in (\ref{RRfields}), it is the
only choice consistent with $5\leftrightarrow 9$ $T$-duality.

At each fixed point the twisted sector includes three NS-NS scalars and
one R-R scalar.  They belong to a single $N=1$ $D=6$ hypermultiplet,
transforming as an $SU(2)_R$ doublet $\Phi$.  Because supersymmetry
remains unbroken, the Higgs mechanism that gives mass to the vector
boson by eating the R-R field must also give mass to the NS-NS
superpartners; let us see how this arises from the $D$-terms.  For any
$U(1)$ gauge symmetry supersymmetry requires that the three one-forms
\begin{equation}
2 I^A \epsilon = i\delta \Phi^* \sigma^A d\Phi - id\Phi^* \sigma^A
\delta \Phi
\end{equation}
be closed.  The $SU(2)_R$-triplet $D$-terms are then defined by $dD^A =
I^A$. For a linearly realized $U(1)$, $\delta \Phi = i\epsilon q
\Phi$ and $D^A = q\Phi^* \sigma^A \Phi$.  For the spontaneously broken
$U(1)$,
\begin{equation}
\delta \Phi = \epsilon v \equiv \epsilon\left[
\begin{array}{c}1\\0\end{array}\right] .
\end{equation}
Parameterizing the doublet by $\Phi = (\phi - i H^A \sigma^A)v$, this
gives $\delta\phi = \epsilon$ and $D^A = H^A$, so that $D^2$ is a mass
term for $H$.

We will consider details of blowing up the orbifold points in section 4.
Let us note briefly here the following. Because of the above remarks,
the D-term contains a term linear in $H^A$ as well as the usual
quadratic terms for charged fields. In particular the D-term for
the 9-brane $U(1)$ contains the sum of all the $H^A_I$. Let us go to a
generic point in the K3 moduli space by giving vevs to all the
geometric moduli of K3. Since the sum of $H^A_I$ is not (generically)
zero, then D-flatness requires the gauge groups to be broken, at
least to $Sp(8)$. We are now on a large smooth K3, and we can try to
identify the
remaining massless
modes in the $\sigma$-model classical limit. $SO(32)$ has an $SU(2)\times
Sp(8)$ maximal subgroup and it is easy to check that if we assume we
have 2 instantons in $SU(2)$ we obtain the correct spectrum.
This corresponds to the following topological fact.
As we will explain in section 4,  the point in  GP moduli space at which
the unbroken gauge group is $U(16)$ (or $SU(16)$ allowing for the
mechanism above) corresponds to a certain $U(1)$ instanton of instanton
number 16, embedded in a very particular way in $Spin(32)/\Z_2$,
on the K3 orbifold.  It can be shown that on a {\it smooth} K3 manifold,
there is no abelian instanton with the given instanton number
and embedding in $Spin(32)/\Z_2$ (this is equivalent to the
fact that upon blowing up, the orbifold singularities  are replaced
by two-spheres whose cohomology classes are not anti-self-dual)
so the $SU(16)$ must be ``Higgsed'' if one turns on twisted sector
modes to blow up the singularities of the K3 orbifold.\footnote{This
is an assertion about K3; the smooth, ALE Eguchi-Hansen
manifold discussed in section four does have a self-dual holomorphic
two-sphere  and
abelian instanton.}  The ``$SU(2)$ instanton of instanton number two''
mentioned above is really an $SO(3)$ instanton; for $SU(2)$ instantons
on a smooth K3 surface, the minimum instanton number is 4, but for an
$SO(3)$ instanton (of non-zero second Stiefel-Whitney class) the minimum
instanton number is 2.  (These last assertions can be understood
using an argument given -- in the $SO(32)$ case -- at the end of section
5.2.)

Repeating the analysis for the 5-branes, and using the ${U(1)}_5$
D-term, we obtain an $Sp(8)$ unbroken 5-brane group after turning on
twisted sector modes, as expected from eight 5-branes on a smooth K3.

\subsection{Anomalies in the General Case}

We now consider the problem in more generality.  Ref. \cite{gimpol}
described models with nine-brane gauge group $U(16)$ and five-brane
gauge group $\prod_I U(m_I) \times \prod_J Sp(n'_J)$.  The five-brane
unitary factors arise from $m_I/2$ five-branes at fixed point $I$, while
the symplectic factors are from 5-branes away from fixed points.  The
maximum rank is achieved when all 8 five-branes sit at fixed points.  We
will focus on this case, since the more general case is obtained by
spontaneous breaking (moving 5-branes off the fixed points).  In
particular, we will consider the $U(1)^{16}$ subgroup of the five-brane
gauge group, coming from open strings with both ends on the same
five-brane.

$T$-duality interchanges five-branes and nine-branes, but the groups
above are not $T$-dual because we have omitted the Wilson lines that are
dual to the five-brane positions \cite{gimpol}.  Let us now rectify
this.  The Wilson lines in the directions $m = 6,7,8,9$ must satisfy
\begin{eqnarray}
[W_m,W_n] = 0, \qquad  M W_m M^{-1} = W_m^{-1},
\end{eqnarray}
{\it i.e.} flatness and the orbifold projection.  Here $M$ is
the matrix in (\ref{chpat}), but it is now regarded as an element of the
group.
We can take a basis in which the $W_m$ are made up of
blocks of two types
\begin{equation}
W_m=\left[ \begin{array}{cccccccc} {\cal R}_2(\theta)&&&&&&&\\
&.&&&&&&\\
&&\pm 1&&&&&\\
&&&.&&&&\\
&&&&{\cal R}^{-1}_2(\theta)&&&\\
&&&&&.&&\\
&&&&&&\pm 1&\\
&&&&&&&.
\end{array} \right]
\label{wils}
\end{equation}
where ${\cal R}_2(\theta)$ is a $2\times 2$ rotation matrix.  The
structure of large and small blocks must be the same for all $m$, by
flatness.  The large $2\times2$ blocks are $T$-dual to the motions of
five-branes in the bulk of the orbifold, while the small blocks are
$T$-dual to half-five-branes fixed at orbifold points.

In parallel with the five-brane discussion, we focus for the remainder
on Wilson lines consisting entirely of small blocks---$T$-dual to all
five-branes on fixed points.  Thus,
\begin{equation}
\label{fixwilson}
W_m = {\rm diag}(w_{m,i}), \quad w_{m,i} = \pm 1, \quad w_{m,i} =
w_{m,i+17}.
\end{equation}
For each fixed Chan-Paton factor $i$, the
$(w_{1,i},w_{2,i},w_{3,i},w_{4,i})$ form one of $2^4$ possible
sequences, which we label $w_{m,\tilde I}$, $\tilde I = 1,\ldots,16$.
The sequences correspond to the fixed points of the $T$-dual theory,
which are Fourier-dual (on $\ZZ_2^4$) to the original fixed points
(after picking a fixed point around which to T-dualize); again, the
latter are labeled $I = 1,\ldots,16$. Note that if $w_{m,i}$ are equal
for different $i$, this is T-dual to having several half-five-branes at
the same fixed point. In any case, the choice (\ref{fixwilson}) gives a
maximal number of $U(1)$'s (more generally, Cartan elements of $U(n)$).

Now let us obtain the spectrum.  The $U(1)^{16}_5 \times U(1)^{16}_9$
gauge transformations are respectively
\begin{equation}
\epsilon = \left[ \begin{array}{cc} 0&\Delta\\-\Delta&0 \end{array}
\right], \qquad
\tilde\epsilon =
\left[ \begin{array}{cc} 0&\tilde\Delta\\-\tilde\Delta&0
\end{array} \right]
\end{equation}
with $\Delta = {\rm diag}(\epsilon_i)$, $\tilde\Delta = {\rm
diag}(\tilde\epsilon_i)$.  These act on the Chan-Paton wavefunctions
$\lambda$ as $\delta\lambda= \epsilon\lambda - \lambda\epsilon$ in the
55-sector, with appropriate tildes in the 99- and 59-sectors.  As noted
in \cite{gimpol}, the 55 scalars have wavefunctions
\begin{equation}
\lambda = \left[ \begin{array}{cc} A_1&A_2\\A_2&-A_1 \end{array}
\right]
\end{equation}
for antisymmetric matrices $A_{1,2}$.  The gauge transformations act
on these as
\begin{equation}
\delta(A_1 + i A_2) = -i\{ \Delta, A_1 + i A_2 \},
\end{equation}
implying charges $\epsilon_i + \epsilon_j$ for $i \neq j$.  The content
is that of two hypermultiplets so we have counted $i \leftrightarrow j$
as independent.  Similarly in the 99-sector we have charges
$\tilde\epsilon_i + \tilde\epsilon_j$ for $i \neq j$.

Now consider the 59-sector with the five-brane at fixed point $I$.  This
is fixed by the operation $T_{2I} R$, where $R$ is the orbifold $\ZZ_2$
and $T_{2I}$ is a lattice translation by twice the coordinate of the
fixed point.  This acts on the nine-brane Chan-Paton factor as $W_{I}
\gamma_{9,R}$ where $W_{I}$ is the Wilson line corresponding to the
translation and $\gamma_{9,R} = M$.\footnote{That is, $T_{2I}$ is a
translation by one lattice unit in some subset of the directions $m$,
and $W_I$ is a product of those $W_m$.  Similarly, $w_{I,j}$, the
$j^{\rm th}$ diagonal element of $W_I$, is a product of the
corresponding elements $w_{m,j}$, and $w_{I,\tilde J}$ is a product of
the corresponding $ w_{m,\tilde J}$ defined below
equation~(\ref{fixwilson}).}

Thus the 59 wavefunctions satisfy\footnote {The dimensions of the $M$'s
on the two sides are in general different, since $\lambda$ need not be
square, but to avoid burdening the notation the same symbol is used.}
\begin{equation}
\lambda = M \lambda M^{-1} W_{I}^{-1}
\end{equation}
with solution
\begin{equation}
\lambda = \left[ \begin{array}{cc} X_1&X_2\\-X_2 W_{I}&X_1 W_{I}
\end{array} \right]\ .
\end{equation}
The gauge transformation is
\begin{equation}
\delta (X_1 + iX_2) = i \Delta (X_1 + iX_2)W_{I} - i (X_1 + iX_2)
 \tilde\Delta .
\end{equation}
The element $(X_1 + iX_2)_{ij}$ thus has charge
$\tilde\epsilon_j-\epsilon_i w_{I,j}$, where $w_{I,j}$ is the $j^{\rm
th}$ diagonal element of $W_I$.

Including the contribution of the hyper and vector multiplets, the
$U(1)$ gauge anomaly is
\begin{eqnarray}
\label{gaugeanomaly}
&&-\sum_I \sum_{i,j\in I} (F_i - F_j)^4
-\sum_{\tilde I} \sum_{i,j\in \tilde I} (\tilde F_i - \tilde F_j)^4
\nonumber\\
&& + \sum_I \sum_{i\neq j\in I} (F_i + F_j)^4
+ \sum_{\tilde I} \sum_{i\neq j\in \tilde I} (\tilde F_i + \tilde F_j)^4
\nonumber\\
&&+\sum_{I,\tilde J} \sum_{i\in I} \sum_{j\in \tilde J}
(F_i - w_{I,\tilde J} \tilde F_j)^4
\\
&=& 6 \sum_{i } F_i^2\,
\sum_{j} \tilde F_j^2
\nonumber\\
&&+ \sum_I \left( 4 {\textstyle\sum_{i \in I}} F_i - {\textstyle
\sum_{\tilde J}} w_{I,\tilde J} {\textstyle\sum_{j \in \tilde J}}
\tilde F_j \right)
\left( 4{\textstyle\sum_{i \in I}} F_i^3 - {\textstyle \sum_{\tilde J}}
w_{I,\tilde J} {\textstyle\sum_{j \in \tilde J}} \tilde F_j^3 \right).
\nonumber\end{eqnarray}
We have used
$\sum_I w_{I,\tilde J} w_{I,\tilde J'} = 16\delta_{\tilde J,\tilde J'}$.
There are also mixed gravitational-$U(1)$ anomalies:
\begin{eqnarray}
&&-3\trr\left( \sum_{i } F_i^2\,+ \sum_{j} \tilde F_j^2\right)
\\
&&-{1\over 16}\trr \sum_I \left( 4 {\textstyle\sum_{i \in I}} F_i -
{\textstyle
\sum_{\tilde J}} w_{I,\tilde J} {\textstyle\sum_{j \in \tilde J}}
\tilde F_j \right)
\left( 4{\textstyle\sum_{i \in I}} F_i - {\textstyle \sum_{\tilde J}}
w_{I,\tilde J} {\textstyle\sum_{j \in \tilde J}} \tilde F_j \right).
\nonumber\end{eqnarray}
These reduce to the anomaly (\ref{anomaly})
when the Wilson lines are removed.

The first term on the right-hand side of eq. (\ref{gaugeanomaly})
factorizes in the usual form and can be canceled by the exchange of the
$B_{\mu\nu}$ field.  The second term can be canceled by exchange of R-R
zero-form fields.  The last line of eq. (3.30) in ref. \cite{gimpol}
gives the tadpole of the R-R twisted sector six-form at fixed point $I$.
Generalizing to include the nine-brane Wilson line, it becomes
\begin{equation}
B_{6,I} \left( 4 {\rm Tr}_I(\gamma_{R,I}) - {\rm Tr} (W_I
\gamma_{R,9}) \right).
\end{equation}
As noted in ref. \cite{gimpol}, the traces vanish automatically once all
other constraints are satisfied.  In fact, this is simply due to
orientation symmetry, under which $B_{6,I}$ is odd and therefore
projected out.  From this calculation we can easily deduce the
additional terms that are needed.  As is clear from the boundary-state
formalism in gauge backgrounds, one gets all 6-forms that can be formed
out of the R-R forms and the gauge field strengths~\cite{bound}.  The
nonvanishing terms are thus
\begin{eqnarray}
\label{jpcounter}
B_{0,I} \left( 4 {\rm Tr}_I(\gamma_{R,I} F^3) - {\rm Tr} (W_I
\gamma_{R,9} \tilde F^3) \right) + B_{4,I} \left( 4 {\rm
Tr}_I(\gamma_{R,I} F) - {\rm Tr} (W_I \gamma_{R,9} \tilde F) \right)&&
\nonumber\\
\propto B_{0,I} \left( 4 {\textstyle\sum_{i \in I}} F^3_i - {\textstyle
\sum_{\tilde J}} w_{I,\tilde J} {\textstyle\sum_{j \in \tilde J}}
\tilde F^3_j \right) + B_{4,I} \left( 4 {\textstyle\sum_{i \in I}} F_i -
{\textstyle
\sum_{\tilde J}} w_{I,\tilde J} {\textstyle\sum_{j \in \tilde J}}
\tilde F_j \right).&&\nonumber\\
\end{eqnarray}
with similar terms involving $\trr$.  These are of just the form required
to cancel the $F\,F^3$ anomaly found in (\ref{gaugeanomaly}).  Note that
$B_{0,I}$ and $B_{4,I}$ are electric and magnetic potentials for the
same field strength.  We must thus carry out a duality transformation to
put the action in local form, schematically
\begin{equation}
B_{0}\wedge F^3 + B_{4}\wedge F \to
(dB_0 + A)\wedge *(dB_0 + A) + B_{0} F^3. \label{gs13}
\end{equation}
The modified gauge transformation $\delta B_0 = -\epsilon$ implied by the
kinetic term gives an $F\,F^3$ anomaly.  We see that the couplings of
the twisted-sector fields are of the precise form needed to cancel the
one-loop anomaly.

The $B_4 \wedge F$ coupling (or equivalently $A\wedge *dB_0$) makes the
would-be anomalous $U(1)$ fields massive as discussed above.  For
example, let the five-branes be distributed half per fixed point to give
the maximum $U(1)^{16}$. At the massless level one can think of
$B_{4,I}$ as a Lagrange multiplier determining the gauge field $F_I$ in
terms of the five-brane fields, so the gauge group is identical to the
nine-brane group, though the actual fields are linear combinations of
the five- and nine-brane fields.  We are considering models with a
rank-32 group, which contains at least two $U(1)$'s (the case considered
in the previous section) and at most 32 (when there is one five-brane at
each $I$ and one nine-brane with each $\tilde I$).  One can see that if
there are 16 or fewer $U(1)$'s, all are broken, while if there are more
than 16, exactly 16 are broken.

\subsection{A Nonperturbative Anomaly}

The models described in the previous section fall into several classes.
The number of half five-branes on each of the 16 fixed points can be odd
or even, and this is fixed because only whole five-branes can move off
the fixed point.  Similarly, the number of Chan-Paton
factors with each $\tilde I$ can be odd or even.
Although these models are all consistent in
perturbation theory, there is a nonperturbative inconsistency in most of
the classes.  Consider transporting a charged field around the fixed
point at $X^m = 0$, from $X^m$ to $-X^m$.  On this nontrivial path the
field picks up the $SO(32)$ transformation $M$.\footnote {The matrix $M$
has been redefined in eq.~(\ref{chpat}) by a factor of $i$ relative to
ref.~\cite{gimpol} so as to lie in the group $SO(32)$.  This is
irrelevant for the tensor representations that appear in perturbation
theory, but necessary to extend the transformation to other
representations.} Now transport a field twice around the fixed point, so
that it comes back to itself times $M^2$.  This path is topologically
trivial so fields must come back to their original values.  Now $M^2 =
-1$, but this is no problem because this is in the $SO(32)$ vector
representation, and there are no fields in this representation.  The
matrix $M$ can be thought of as a rotation by $\frac{1}{2}\pi$ in each
of the 16 planes $(k,k+16)$.  It follows that $M^2$ is $+1$ in one
spinor representation and $-1$ in the other.  This is precisely the
spectrum given by the GSO projection, so the holonomy is consistent.

Now consider the fixed point $I$.  This is fixed by $T_{2I}R$, so the
holonomy is $W_I M$ and the consistency condition is $W_I M W_I M = 1$
in the relevant representations.  By construction, the Wilson
line~(\ref{wils}) satisfies $M W_I M^{-1} = W_I^{-1}$ and so $W_I M W_I
M = M^2$ in the vector representation.  In spinor representations it is
not hard to see that the large blocks (being connected to the identity)
have no net effect.  Consider a small block with $-1$'s in rows $k$ and
$k+16$.  One can think of this as a rotation by $\pi$ in the plane
$(k,k+16)$.  This commutes with $M$ and squares to a $2\pi$ rotation.
Thus it is trivial in the vector representation but $-1$ in both
spinors.  If there are an odd number of such blocks, then $W_I M W_I M =
-M^2$ in the spinor representation and there is no spinor representation
that can be consistently defined at {\it both} fixed points.
Consistency of the model thus requires that for all $I$ the number of
small $-1$ blocks is even.  The $T$-dual statement for the five-branes
is this: consider any 8 fixed points lying in a 3-plane (such as the
3-plane $x^6= 0$).  The total number of five-branes on these fixed
points must be integer.

Since the spinor representations appear only as D-branes, this is a
nonperturbative inconsistency.  In section~5 we will see that it has a
simple topological interpretation, and we will actually find a slightly
stronger condition: consider any 4 fixed points lying in a 2-plane (such
as the 2-plane $x^6=x^7= 0$).  The total number of five-branes on these
fixed points must be integer.

Taking into account the nonperturbative constraint, there appear
to be six connected sets of GP orbifolds.  There can be zero
half-five-branes, 8 half-five-branes in a 3-plane, or
16 half-five-branes, times the corresponding three sectors of
Wilson lines, giving nine sectors that are reduced to six by
T-duality.
 We will see in section~5 that all of them are connected
via smooth $K3$'s.

The topological interpretation of the non-perturbative
  inconsistency that we have described really has to do
with the second Stiefel-Whitney class $w_2$ of the $SO(32)$ bundle.
In the above discussion, the key point was that
a reflection $R$ of $\R^4$ and a translation $T_I$ in the $I^{th}$
direction obey geometrically
$$RT_I=T_I^{-1}R.$$
In dividing by the group $G$ generated by $R$ and $T_I$ to make a K3
orbifold from $\T^4$, we represented $T$ and $T_I$ in the gauge group
$SO(32)$ by matrices $M$ and $W_I$,
in such a way that the desired relation
$$MW_I=W_I^{-1}M,$$
is satisfied in $SO(32)$ but not after lifting to $Spin(32)/\Z_2$.
In such a situation, if $G$  acted freely on $\R^4$, then after dividing
$\R^4$ by $G$, one would get a smooth manifold $\R^4/G$
with a flat $SO(32)$ bundle of non-zero $w_2$.  Here, we are
dealing with a somewhat more abstract string-theoretic
version of $w_2$, as the $G$ action is not free.

\sect{Duality$_1$: Type I and Heterotic $SO(32)$}

In ten dimensions the $SO(32)$ Type~I and heterotic strings are dual
with a specific transformation of the metric and other
fields~\cite{witusc}, and so they remain equivalent when compactified on
corresponding smooth manifolds.  Finding the dual of an orbifold
compactification is less straightforward.  In the limit of a large
orbifold, one can use the adiabatic principle~\cite{adi}: the dual
theory looks geometrically like an orbifold away from the fixed point.
Also, by transporting charged fields around a fixed point, it follows
that the gauge holonomy (here $M$) is the same in both theories. But it
need not be that a free orbifold CFT in one theory maps to a free
orbifold theory in the other---there might, for example, be
twisted-state backgrounds.  This arose in ref.~\cite{gimpol}, where the
heterotic orbifold with spin and gauge connections equal did not have a
free-CFT Type I dual. In the present case, we can see immediately that
most of the Type I models do not have a free-CFT dual.  The point is
that the antisymmetric tensor 6-form charge is canceled locally only
for one set of models, the $U(1)^{16}_5$ models with a half-five-brane
at each fixed point.  For the other models there is a local charge,
hence a field strength and a dilaton gradient.  This is a higher order
(disk) effect in the Type I CFT, but tree level in the heterotic dual.

In section~3.1 we first consider some generalities that do not depend on
such details, specifically the relation between Type I $T$-duality and
heterotic weak/strong duality~\cite{duff,dmw}.  In section~3.2 we study
those models which do have free-CFT heterotic duals.

\subsection{Type I - Heterotic Duality}

The quadratic-quadratic anomaly polynomial is
\begin{equation}
(\trr - 2 \tr \tilde F^2)(\trr - 2 \tr F^2),
\end{equation}
where (as before) a tilde is used for the nine-brane fields, though now
the trace is in the fundamental representation of $U(m)$ rather than the
Chan-Paton representation.  In the heterotic string description, either
$F$ or $\tilde F$ is identified as a perturbative gauge field at level
one, whereas the other is identified as a nonperturbative gauge
field~\cite{sag,dmw}.  It follows that $T$-duality, which interchanges
$F$ and $\tilde F$ in the Type I description, maps to strong/weak
duality~\cite{duff,dmw} in the heterotic description.

To develop this further, let us identify some of the scalars that appear
in heterotic and Type I compactifications to six dimensions.  The main
point is to organize them in terms of supersymmetry multiplets.  This
gives a preferred (and most convenient) basis.  Field redefinitions are
constrained by gauge invariance and supersymmetry.  Since the tensor
multiplet includes a gauge field, the scalar $\phi$ in the multiplet
cannot be redefined by an infinitesimal transformation.  The only
freedom is to multiply the tensor multiplet by minus one, which acts on
the bosons in the multiplet as
\begin{equation}
\phi \rightarrow -\phi , \qquad B^- \rightarrow -B^- .
\label{dualtrans}
\end{equation}
This transformation on $B^-$ is equivalent to a duality transformation on
the two-form $B$.  In general, this duality transformation is not a
symmetry of the Lagrangian.  It maps it to another Lagrangian.

We are now going to identify a particular scalar field $\phi$ (and one
hypermultiplet) in two cases:
\begin{enumerate}
\item The heterotic string on K3.  The hypermultiplets
include the moduli of K3, including the radius $r_h$.  Since supersymmetry
transformations act simply on the background fields in the world-sheet
sigma model, $r_h$ is the radius in the heterotic string metric.
The ten dimensional dilaton $D_h$ affects the scalar $\phi$ in the
tensor multiplet---the latter is a function of $D_h$ and $r_h$.  To
identify $\phi$, consider the ten dimensional Lagrangian expressed in
terms of the heterotic string metric.  One of the terms there is
$e^{-2D_h} H \cdot H$.  Therefore, the coefficient of $H \cdot H$ in the
six dimensional Lagrangian in the string metric is $r_h^4e^{-2D_h}$.
Since this term is Weyl invariant, this is also the answer in the
Einstein metric, and hence
\begin{equation}
r_h^4e^{-2D_h} = e^{-2\phi}.
\end{equation}
\item The Type I theory on K3.  The scalar $\phi$ is again a
function of two scalars: the ten dimensional dilaton $D_I$ and the radius
of K3, $r_I$, in the Type I string metric.  To identify $\phi$, consider
the ten dimensional Lagrangian expressed in terms of the Type I string
metric.  Since $B$ is a RR field, the $H \cdot H$ term in that
Lagrangian is not multiplied by an exponential of $D_I$.  Therefore, the
coefficient of $H \cdot H$ in the six dimensional Lagrangian in the
string metric is $r_I^4$.  Since this term is Weyl invariant, this is
also the answer in the Einstein metric, and hence
\begin{equation}
r_I^4= e^{-2\phi}.
\end{equation}

Note that $\phi$ is independent of the ten dimensional dilaton $D_I$.
The same conclusion can also be reached by studying the action of
spacetime supersymmetry transformations on the world-sheet sigma model.
It acts on the moduli without mixing with $D_I$ which multiplies the two
dimensional curvature term.  Therefore, $r_I$ should appear as a field
in a separate multiplet.  The ten dimensional dilaton $D_I$ affects the
hypermultiplets.  A straightforward dimensional reduction shows that the
appropriate combination is $e^{-2D_I}r_I^4$. \end{enumerate}

In fact, given the identification of the fields in the heterotic
compactification we could have derived it in the Type I compactification
using the change of variables in the ten dimensional Lagrangian used in
heterotic/Type I duality (for this purpose we need only the change of
variables, not the assumption of complete duality)
\begin{equation}
r_I^2=r_h^2e^{-D_h}, \qquad
D_I=-D_h,
\end{equation}
which identifies the two fields as
\begin{equation}
e^{-2\phi}=r_I^4=r_h^4e^{-2D_h}, \qquad r_I^4 e^{-2D_I}=r_h^4 ,
\end{equation}
i.e.\ as the Type I radius and the heterotic radius.

The duality transformation~(\ref{dualtrans}) in the Type I variables
inverts the Type I radius $r_I$ while holding $r_h= r_I e^{-D_I/2} $
fixed.  This is precisely the action of $T$ duality in this theory.  So
Type~I $T$-duality is the image of heterotic weak/strong duality, which
inverts the six-dimensional heterotic coupling $e^{\phi}$. Note that the
fact that $r_h$ is held fixed and therefore $D_I$ transforms is standard
in T duality, which keeps the Newton constant $G_N=r_h^4= r_I^4 e^{-2D_I} $
fixed.


Heterotic duality gives two different perturbative heterotic limits of
the GP theory.  The latter should therefore have two different heterotic
string solitons, one carrying the current algebra of the nine-brane
gauge group and one that of the five-brane gauge group, and these should
be interchanged by $T$-duality.  The first is just the Dirichlet
one-brane, extended in a non-compact direction.  Its $T$-dual is the
Dirichlet five-brane, wrapped in the compact directions and extended in
one non-compact direction.  The analysis in GP can be extended to
include these objects.  The Chan-Paton algebra gives
\begin{equation}
\gamma_{\Omega,1} = \gamma_{\Omega R,5'} = I,
\qquad
\gamma_{R,1} = \gamma_{\Omega R,1} = \gamma_{R,5'} = \gamma_{\Omega,1} =
M.
\end{equation}
There are no constraints from divergences, because the R-R flux is free
to spread in the non-compact directions.  A prime is used to distinguish
the heterotic five$'$-branes, which are wrapped on the $K3$ and
localized in four non-compact directions, from the GP five-branes, which
are extended in all the non-compact directions.  Since $M$ is
even-dimensional, the minimal five$'$-brane thus has a two-valued
Chan-Paton index as found in refs.~\cite{sminst,gimpol}.  The minimal
one-brane also has a two-valued Chan-Paton index; this simply labels the
one-brane, at a given point on the compact space, and its image under
$R$.

Quantization of the one-brane is precisely as in ref.~\cite{polwit}, as
long as the one-brane is not coincident with a fixed point or
five-brane.  One finds right- and left-moving oscillations, in both the
noncompact and compact directions, together with right-moving
Green-Schwarz superpartners.  There are also real left-moving fermions
>from the 19- open strings.  These carry the 32-valued nine-brane index.
They generate an $SO(32)$ current algebra, since the Wilson lines and
orbifold projections do not affect the local structure on the
macroscopic one-brane.  There is no current algebra from the 15-strings
because these are massive, being stretched.

When the one-brane sits at a fixed point there are additional massless
degrees of freedom from strings stretched between the one-brane and its
image.  These are found to be a $U(1)$ gauge field on the one-brane;
also, the oscillations in the compact directions are enlarged to a
complex field carrying the $U(1)$.  Their right-moving superpartners
also become complex, and a real left-moving fermion appears.  Similarly,
when the one-brane is coincident with a five-brane there are massless
right- and left-moving fields in the 15 sector.  These carry the
32-valued five-brane index, and the NS and R strings transform
respectively as spinors $\bf 2$ under the noncompact and compact $SO(4)$
tangent groups.  Quantum fluctuations of a one-dimensional object will
take it away from these special points, but it may be possible to fix it
by turning on vev's of the non-generic massless fields, and in this way
find new strings.

For the five$'$-brane everything is the same by $T$-duality, with the
five-branes and nine-branes interchanged.  In particular, the transverse
position of the one-brane in the compact direction is related by
$T$-duality to a $U(1)$ gauge field living on the five$'$-brane tangent
to the non-compact directions.

It is straightforward to calculate the tension of the two strings.  In
the Type I theory the electric heterotic string (the ten-dimensional
one-brane) has tension
\begin{equation}
T_e={1 \over \lambda_{\rm I}}=e^{-D_{\rm I}}= {r_h^2 \over r_{\rm I}^2}.
\end{equation}
The magnetic heterotic string (the ten-dimensional five-brane) wraps
around K3 and therefore its tension is proportional to the volume of K3
\begin{equation}
T_m={r_{\rm I}^4\over \lambda_{\rm I}}= r_{\rm I}^4 e^{-D_{\rm I}}= r_h^2
r_{\rm I}^2.
\end{equation}
Clearly, they are exchanged under $r_{\rm I} \rightarrow 1/r_{\rm I}$
holding $r_h$ fixed, which is our duality transformation.

\subsection{An Orbifold Dual}

The only GP theories for which the $H$ charge is canceled locally, and
which therefore might map to free theories on the heterotic side, are
those with the eight five-branes distributed half per fixed point.
Indeed, as we will now describe, there is a consistent $\T^4/\Z_2$
orbifold of the heterotic string, constructed by embedding $\Z_2$ in
$Spin(32)/\Z_2$ using the matrix $M$; its massless spectrum matches that
of a GP model with half a five-brane at each fixed point.

Let us focus on the case with no Wilson lines.  First we describe
what one sees on the Type I side.  The Type I
gauge group, taking into account the results of section~2, is
$SU(16)\times U(1)$.  Turning to the hyper multiplets, the 99-sector
contributes two antisymmetric {\bf 120}'s, and the 59-sector contributes
a ${\bf 16}$ for each fixed point.  The $\bf 16$ at fixed point $I$
couples to the $U(1)$ linear combination $A_I + \tilde A$.  It follows
{}from section~2 that the massless gauge bosons have $A_I = -4
\tilde A$, so the $\bf 16$ couples to $-3
\tilde A$; the
$U(1)$ charge of each $\bf 16$ is $-3$, where that of each $\bf 120$ is
$+2$.

Now we consider the heterotic string orbifold in which dividing
by the reflection $R$ is accompanied by a transformation $M$ in the
gauge group.
In the untwisted sector, the projection to $R$-invariant states
reduces the gauge group from $SO(32)$ to $U(16)$, while the
surviving hypermultiplets (coming from the components of the gauge
field tangent to $\R^4$) are two $\bf 120$'s.  Note that the
holonomy is order 4 in the twisted sector, $M^2 = -1$ being the current
algebra GSO projection.  The current algebra fermions are thus moded in
integers+$\frac{1}{4}$ and the zero point energy on the left side is
\begin{equation}
-\frac{32}{192} -\frac{4}{24} +\frac{4}{48} = -\frac{1}{4}.
\end{equation}
This is level-matched, the massless level having one $\lambda^a_{-1/4}$
excitation with $a$ an $SU(16)$ {\bf 16} index, showing the consistency
of the theory.  The zero point shift of
the $U(1)$ charge is $-\frac{16}{4}$, giving the net $-3$ as above.
Similarly the sector twisted by $M^3$ gives the $\overline {\bf 16}$
with charge $+3$.  So the twisted sector massless states agree with
what is found in the Type I description.

\sect{Topology Of The GP Model}

The GP models are, of course,  Type I orbifolds with target
space $\R^6\times \T^4/\Z_2$.  One expects intuitively that the twisted
sector fields supported at the $\Z_2 $ fixed points should include
blowing up modes associated with a deformation from $\T^4/\Z_2$ to a
smooth K3 surface; indeed, the closed string spectrum found
in~\cite{gimpol} contains appropriate twisted sector fields to do the
job.  There are, however, a variety of puzzles about the connection of
the GP model with K3 compactifications that we wish to unravel here.

To explain the issues, we may start by noting the following.  A smooth
K3 compactification needs a vacuum $SO(32)$ gauge bundle with instanton
number 24, to make it possible to obey the familiar equation $dH= \tr
R\wedge R-\tr F\wedge F$.  The GP model does not have any explicit
instantons in the vacuum.  Rather than instantons, the model has
five-branes; there are eight five-branes, at arbitrary positions on
$\T^4/\Z_2$.\footnote{As noted in footnote~1, these eight five-branes
come from 32 Chan-Paton indices, counting the images under the
orbifold $\Z_2$ and world-sheet parity $\Omega$.} Since a Type I
five-brane is equivalent to the small size limit of an instanton
\cite{sminst}, eight instantons are implicit in the five-branes of the
GP model.  Since 24 are expected, $24-8=16$ seem to be missing.

As there are 16 $\Z_2$ orbifold singularities in $\T^4/\Z_2$, one might
intuitively think that one instanton is ``hiding'' at each singularity,
and that blowing up one of these singularities will bring an instanton
out into the open.  Understanding this will be our first goal.

The construction in~\cite{gimpol}, as we have explained above,
 involves a twist operator that acts on
the $SO(32)$ Chan-Paton label by multiplication by a $32\times 32$
matrix which in $16\times 16$ blocks looks like
\begin{equation}
M=\left(\begin{array}{cc} 0 & I \cr -I & 0\end{array}\right).
\label{multby}
\end{equation}
As noted earlier, this matrix obeys not $M^2=1$, as one might expect in
constructing a $\Z_2$ orbifold, but $M^2=-1$.  Thus, this orbifold would
not be possible if the gauge group of the Type I superstring were really
$SO(32)$.  Its viability depends on the fact that the gauge group is
really $Spin(32)/\Z_2$.  The $\Z_2$ in question is generated by an
element $w$ of the center of $Spin(32)$ that acts as $-1$ on the 32
dimensional vector representation, $-1$ on one spinor, of, say, negative
chirality, and $+1$ on the other spinor.  Only representations with $
w=1$ are present in the $Spin(32)/\Z_2$ heterotic string, or
equivalently, in the Type I superstring.  The matrix $M$ obeys $M^2=w$,
and in the Type I or $Spin(32)/\Z_2$ heterotic theory, this is
equivalent to $M^2=1$.

Thus, a $\Z_2$ orbifold such as this one is possible.  But its existence
depends on the fact that the gauge group is $Spin(32)/\Z_2$ rather than
$SO(32)$, and we will have to use this fact in comparing the model to
what can be seen geometrically.

We begin with some remarks about the difference between $SO(32)$ and
$Spin(32)/\Z_2$ vector bundles on a manifold $X$.  It is possible to
have a $Spin(32)/\Z_2$ vector bundle that is not associated with any
$SO(32)$ vector bundle.  This can be achieved if on some two-cycle
$S\subset X$, Dirac quantization is obeyed for the adjoint
representation, and the positive chirality spinor, but not for the
vector or negative chirality spinor.  If for some $Spin(32)/\Z_2$
bundle, precisely the representations that are present for
$Spin(32)/\Z_2$ obey Dirac quantization, then this bundle cannot be
derived from an $SO(32)$ bundle.  For an explicit example, suppose that
the gauge field lives in an abelian subgroup of $Spin(32)/\Z_2$,
generated by a matrix $Q$ which is the sum of 16 copies of
\begin{equation}
\left( \begin{array}{cc}
0 & 1 \cr -1 & 0
\end{array} \right).  \label{lefgam}
\end{equation}
Suppose moreover that the integrated magnetic flux is $\pi$ -- that is
precisely one-half of a Dirac quantum.  Then Dirac quantization is
violated for the vector, but is obeyed for the adjoint or the positive
chirality spinor (for which the sum of the 16 $U(1)$ charges is even).
This is the basic example of a $Spin(32)/\Z_2$ bundle that is not
associated with an $SO(32)$ bundle.  Note that the fact that $F/2\pi$
has a half-integral number of Dirac quanta for {\it every} element of
the vector representation is essential in ensuring that Dirac
quantization is obeyed for the adjoint and positive chirality spinor
representations.  This is the reason for using the embedding via $Q$.

Given a $Spin(32)/\Z_2$ bundle $E$, one can define a mod two cohomology
class $\tilde w_2(E)$, which assigns the value $+1$ to a two-cycle on
which Dirac quantization is obeyed for the vector representation and
$-1$ to a two-cycle on which it is not obeyed.  Thus $\tilde w_2(E)\in
H^2(X,\Z_2)$ measures the obstruction to associating $E$ with an
$SO(32)$ bundle.  The notation $\tilde w_2$ is motivated by the fact
that the obstruction to deriving a $Spin(n)$ bundle from an $SO(n)$
bundle $F$ is conventionally called $w_2(F)$ ($w_2$ is the second
Stiefel-Whitney class).  $\tilde w_2$ is quite analogous to $w_2$; in
fact, for $n=8$, $Spin(8)$ triality exchanges them.  One might describe
$\tilde w_2$ as the obstruction to a bundle having ``vector structure,''
just as $w_2$ is the obstruction to ``spin structure.''

\subsection{Blow-Up of a $\Z_2$ Orbifold Singularity}

Now we want to consider the behavior of a $Spin(32)/\Z_2$ bundle near
one of the $\Z_2$ orbifold singularities $P\in \T^4/\Z_2$.  If $P$ is
blown up, one gets a two-sphere $S$, of self-intersection number $S\cdot
S =-2$.  The structure near $S$ looks like the Eguchi-Hansen space,
which is an ALE hyper-Kahler manifold $X$ with fundamental group at
infinity $\Z_2$.  We will call the fundamental group at infinity $\tilde
\pi_1(X)$.  We want to guess what kind of gauge theory on $X$ one gets
as a local description near $S$ after just slightly blowing up the
singularities of the GP model.

First of all, the Eguchi-Hansen space admits a $U(1)$ instanton field
$A$ whose field strength $F=dA$ vanishes at infinity.  If we think of
$S$ as the complex manifold ${\bf P}^1$, then $X$ can be regarded as a
complex line bundle over $S $; in fact, $X$ is the total space of the
line bundle ${\cal O}(-2)$.  This means that if $Y$ is a fiber of $X\to
S$, then
\begin{equation}
{\int_S{F\over 2\pi}=-2\int_Y{F\over 2\pi}.}
\label{kikj}
\end{equation}
We want to eventually embed the $U(1)$ instanton in $Spin(32)/\Z_2$,
using the matrix $Q$, in such a way that Dirac quantization is obeyed
for the adjoint or positive chirality spinor but not for the vector.
With this in mind, we normalize $F$ so that
\begin{equation}
\int_S{F\over 2\pi}={1\over 2},
\label{hujk}
\end{equation}
and hence
\begin{equation}
\int_Y{F\over 2\pi}=-{1\over 4}.
\label{jujik}
\end{equation}
Therefore
\begin{equation}
\int_X {F\wedge F\over 16\pi^2}=-{1\over 32}.
\label{kujik}
\end{equation}
(In doing the integral, one can, using (\ref{jujik}), think of one factor
of $F/2\pi$ as $-1/4$ of a delta function supported on $S$, after which
the integral over $S$ is done using (\ref{hujk}).)

If now this gauge field is embedded in $Spin(32)/\Z_2$ via the embedding
$Q$ of the $U(1)$ Lie algebra into $SO(32)$ (that is, using the sum of
sixteen copies of (\ref{lefgam})), then the $SO(32)$ instanton number
becomes
\begin{equation}
\int_X{\tr F\wedge F\over 16\pi^2} =1.
\label{nujik}
\end{equation}
Thus, this  is an instanton of instanton  number one that obeys
Dirac quantization for the adjoint or the spinor, but not for the
vector.  It is an instanton without ``vector structure.''

Because $F$ is square-integrable, this instanton approaches a flat
connection at infinity.  We can determine which flat connection
it is.  The generator of the fundamental group at infinity
$\tilde \pi_1(X)$ is simply a large  circle
at infinity in $Y$.  The monodromy $W$ of the connection
around this circle is
simply $\exp\int_YF$, and with the embedding (\ref{lefgam}) and the
factor of $-1/4$ in (\ref{jujik}), this is equivalent to
$$
W=M,
$$
with $M$ the twisting matrix (\ref{multby}) used in~\cite{gimpol}.

The gauge field that is related -- upon slight blowing up -- to the GP
model must have monodromy $M$ around the generator of $\tilde\pi_1(X)$,
since in dividing by the $\Z_2$ that creates this cycle, the Chan-Paton
factors were multiplied by the matrix $M$.  We also expect this gauge
field to have instanton number one, since as we explained above, in the
GP construction there seems to be one ``missing'' instanton buried in
each fixed point.  Moreover, it is very natural to suspect that this
gauge field must commute with a $U(16)$ subgroup of $SO(32)$, because GP
models get an unbroken $U(16)$ gauge symmetry from the nine-branes.
(The $U(16)$ is broken to $SU(16)$ by quantum corrections discussed in
section~2.)  The instanton we have constructed does indeed break
$SO(32)$ to $U(16)$, because $U(16)$ is the subgroup of $SO(32)$ that
commutes with $Q$.  Moreover, for a $U(16)$-invariant instanton that
admits spinors but not vectors, the minimum instanton number is one,
since in (\ref{hujk}) we used the smallest half-integer.  (From the
analysis below of the dimension of instanton moduli space, it will be
clear that an instanton $E$ on $X$ with $\tilde w_2(E)\not= 0$ has
instanton number at least one even if one does not assume unbroken
$U(16)$.)  The field we have constructed is clearly the unique
$U(16)$-invariant instanton with instanton number one and monodromy $M$
at infinity, and moreover is overdetermined by those properties.  We
regard these facts as compelling evidence that this is the gauge field
related, after slight blow-up, to the structure of the GP model near the
orbifold singularities.

\subsection{Dimension Of The Moduli Space}

To probe somewhat more deeply, we will need to understand some facts
about instanton moduli spaces both on the non-compact hyper-Kahler
manifold $X$ and on a compact K3 manifold.  In general, with a simple
gauge group $G$ on a compact four-manifold $Y$ without boundary, the
index formula for the dimension of instanton moduli space for instanton
number $k$ is
\begin{equation}
{\rm dim}\,{\cal M}_k=4hk -{\rm dim}\,G\,\left(b_0-b_1
+b_2^+\right),
\label{yurry}
\end{equation}
where $h$ is the dual Coxeter number of $G$, $b_0$ and $b_1$ are the
dimensions of the spaces of harmonic zero-forms and one-forms on $Y$,
and $b_2^+$ is the dimension of the space of self-dual harmonic
two-forms on $Y$.  On K3, $ b_0=1$, $b_1=0$, and $b_2^+=3$, so the
formula becomes
\begin{equation}
{\rm dim}\,{\cal M}_k=4hk-4{\rm dim}\,G.
\label{purry}
\end{equation}
Actually, the formulas (\ref{yurry}) and (\ref{purry}) only coincide
with the actual dimension of instanton moduli space if the generic
instanton number $k$ field completely breaks the gauge symmetry;
this will be so if $k$ is large enough.

Note that if $\pi_1(G)\not=0$, giving the instanton number $k$ does not
uniquely fix the topological class of the instanton; one will also meet
two-dimensional characteristic classes such as $w_2$ and $\tilde w_2$.
(If $G$ is not connected, one also meets one-dimensional characteristic
classes.)  These do not, however, appear in the index formula
(\ref{yurry}) (except indirectly via the fact that $k$ is sometimes
shifted from integral values when classes such as $w_2$ are present).

Now if one wants to consider not a compact manifold $Y$ but an ALE
hyper-Kahler manifold $X$, there are a few modifications in the formula.
The moduli problem we want is one in which the instanton is required to
be flat at infinity.  Also, we do not want to divide by global gauge
transformations at infinity; this has the happy consequence that moduli
space always has the dimension suggested by the index formula, since
there are no constant gauge transformations to worry about (and the
relevant $H^2$ group can likewise be shown to vanish using the
fact that the metric is hyper-Kahler).

Another change is that $k$ might not be an integer (even when
$\pi_1(G)=0$).  In fact, in addition to specifying $k$ (and classes such
as $w_2$), a component of instanton moduli space is labeled by the
choice of a flat connection at infinity, or equivalently the choice of a
representation $\rho$ (in $G$) of the fundamental group at infinity
$\tilde \pi_1(X)$.  The values of $k$ are shifted from integers by an
amount equal to the Chern-Simons invariant of the flat connection.  This
has the intuitively expected consequence that if the fundamental group
at infinity is $\Z_n$ (corresponding to the blow-up of a $\Z_n$ orbifold
singularity), then $k$ is not necessarily an integer but takes values in
$\Z/n$.  Note, though, that (for appropriate $G$) several choices of
$\rho$ may give the same shift in $k$ (we give examples later), so
specifying $k$ does not determine the problem.

Finally, and crucially in what follows, the index formula on an ALE
space is not obtained simply by shifting $k$ as needed.  There is a
crucial contribution involving the eta invariant (for the operator
$d+d^*$ restricted to self-dual forms) of the flat connection $\rho$ at
infinity.  This contribution depends only on $\rho$ (and not on the
instanton number or other characteristic classes); in fact, it only
depends on how the adjoint representation of $G$ transforms under
$\rho$.  Further, the quantities $b_0,b_1,$ and $b_2^+$ cannot simply be
replaced by their ${\bf L}^2$ counterparts (which in fact vanish).  One
must go back to the index theorem, and use the $R^2$ curvature integral
which on a compact manifold would equal $b_0-b_1+b_2^+$, or equivalently
the eta invariant of the trivial flat connection at infinity.

In this paper, the only ALE space that we will consider in detail is the
Eguchi-Hansen manifold $X$, with fundamental group at infinity $\Z_2$.
The representation $\rho$ just corresponds to the choice of an element
$x\in G$ with $x^2=1$.  The eta invariant is a linear combination of the
numbers $n_+$ and $n_-$ of generators of $G$ that are even or odd under
$x$; of course, $n_++n_-={\rm dim}\,G$.  The $R^2$ curvature integral
gives a contribution proportional to ${\rm dim}\,G$.  So the terms
mentioned in the last paragraph are linear combinations of $n_+$ and
$n_-$.  The coefficient of $n_+$ is actually zero, since for abelian $G$
the moduli space has dimension zero. The coefficient of $n_-$ is
actually such that the dimension of moduli space is
\begin{equation}
{\rm dim}{\cal M}_k=4hk -{1\over 2}n_-.
\label{kippo}
\end{equation}

As an example, take $G=Spin(32)/\Z_2$, and set $x$ equal to the matrix
$M$ that appeared earlier.  As $M$ breaks $Spin(32)$ to $U(16)$, and
$\dim\,Spin(32)=496$, $\dim\,U(16)=256$, we have $n_-=496-256=240$.
Also, for $Spin(32)$, $h=30$.  The formula thus becomes in this case
\begin{equation}
{\rm dim}{\cal M}_k=120(k-1).
\label{ippo}
\end{equation}
So the $k=1$ instanton that we constructed earlier with monodromy $M$ at
infinity has no moduli.  Indeed, none were manifest in the construction
of this instanton (and it is not hard to prove directly that there are
none).  Moreover, the spectrum of the GP model, when all five-branes are
safely away from the orbifold singularities, contains no massless
$U(16)$ non-singlets that would be naturally interpreted as moduli of
the instanton gauge bundle at the singularity.  So the fact that in this
case ${\rm dim}{\cal M}_1=0$ is further evidence that the instanton we
constructed is related to the gauge bundle of the GP model.

\sect{Duality$_2$: Heterotic $SO(32)$ and $E_8 \times E_8$}

\subsection{Instantons On The ALE Space}

For further illustration of these idea, we will
need to understand
the various types of $E_8$ or $Spin(32)/\Z_2$ instantons on
the ALE space with fundamental group $\Z_2$ at infinity.  For a recent
discussion on instantons on ALE spaces from the point of view of string
theory see~ \cite{douglasmoore}.

We will need to classify $\Z_2$ subgroups of $Spin(32)/\Z_2$ and $E_8$,
since the monodromy of the instanton
at infinity generates such a subgroup.
First, begin with $Spin(32)/\Z_2$.  There are two types of elements of
order two in $Spin(32)/\Z_2$: those that would square to one in
$SO(32)$, and those that would square to $-1$ in $SO(32)$.  A basic
topological fact is that the monodromy at infinity on
the ALE space is of the second kind
-- it squares to $-1$ in $SO(32)$ -- if and only if the bundle does not
have vector structure, that is $\tilde w_2(E)\not= 0$.  One might
suspect this intuitively, and a proof can go as follows.  The region $T$
at infinity in the Eguchi-Hansen space is homotopic to a circle bundle
over the two-sphere $S$; this circle bundle has Euler class $-2$
(because $S\cdot S=-2$).  Since the Euler class of the bundle reduces to
0 mod 2, the spectral sequence (for the fibration $T\to S$) that
computes the mod 2 cohomology of $T$ is trivial, and the pullback
$H^2(S,\Z_2)\to H^2(T,\Z_2)$ is an isomorphism.  So the bundle lacks
vector structure when restricted to $T$ if and only if it lacks vector
structure when restricted to $S$.  A flat bundle at infinity with
monodromy that squares to one in $SO(32)$ obviously corresponds to a
bundle with vector structure at infinity, and from the special case we
examined of an instanton without vector structure on $S$
with monodromy $M$ at
infinity, it is clear that monodromy that squares to $-1$ corresponds to
lack of vector structure at infinity.  So in short, the monodromy at
infinity
squares to $-1$ in $SO(32)$ if and only if $\tilde w_2(E)\not= 0$.

Now to classify the $\Z_2$ subgroups, consider first the case of a
bundle with vector structure where we are dealing with $\Z_2$ subgroups
of $SO(32)$.  Such a group is generated by a matrix that we can take to
be
\begin{equation}
{x={\rm diag}(-1,-1,\dots,-1,1,1,\dots,1)}
\label{hullo}
\end{equation}
with $p$ eigenvalues $-1$ and $32-p$ eigenvalues $1$.  For this to be in
$SO(32)$ rather than $O(32)$, $p$ must be even. Requiring $x$ to be of
order $2$ (and not order 4) when lifted to $Spin(32)$, $p$ must be
divisible by four; after dividing by $\Z_2$ to get $Spin(32)/\Z_2$, we
can identify $x$ with $-x$ and so take $p\leq 16$.  The non-trivial
cases are thus $p=4,8,12$, and $16$.  This gives four choices of $\Z_2$
subgroup.  Of course $n_-=p(32-p)$.

We now want to show that with such monodromy at infinity, the instanton
number is $k=p/8$ mod $\Z$ (so that integer $k$ corresponds to the three
cases $p=0,8,16$ and half-integer $k$ corresponds to the two cases
$p=4,12$).  One method to do this is to simply exhibit a special case of
an instanton with that instanton number modulo $\Z$.  Take the gauge
group to be $SO(4)$, and ask for the monodromy at infinity to be $-1$.
With $SO(4)$ regarded as $(SU(2)\times SU(2))/\Z_2$, take a standard
$SU(2)$ one-instanton solution on $\R^4$, centered at the origin, and
embedded in one of the $SU(2)$ factors of $SO(4)$.  Such a field is
invariant under the $\Z_2$ symmetry $x^i\to -x^i$ of $\R^4$, and
descends to an instanton number $1/2$ field on $\R^4/\Z_2$ with
monodromy at infinity $-1\in SO(4)$.\footnote{The following facts, which
make this clear, may be familiar.  The region at infinity in $\R^4$ is
homotopic to $\S^3$, which is isomorphic to $SU(2)$.  The $SU(2)$
one-instanton is asymptotic at infinity to a pure gauge $A=dg\cdot
g^{-1}$, where in a suitable gauge $g$ is the ``identity map'' from the
$\S^3$.  Therefore, under the $\Z_2$ transformation of $\R^4$ or $\S^3$,
which acts by ``multiplication by $-1$,'' one has $g\to -g$, making
clear that the monodromy at infinity, after dividing by this $\Z_2$, is
the element $-1$ of $SU(2)$ or equivalently of $SO(4)$.} Thus if the
monodromy at infinity has $4$ eigenvalues $-1$, the instanton number is
$1/2$ modulo $\Z$.  With $4n$ eigenvalues $-1$ at infinity, one can take
$n$ copies of the half-instanton just described in commuting $SU(2)$
subgroups of $SO(4n)$, giving instanton number $n/2$ modulo $\Z$, in
agreement with the claim above.  The same method, applied to an $SO(16)$
subgroup of $E_8$, can be used to justify the claims made presently for
$E_8$.

In this analysis of $\Z_2$ subgroups of $SO(32)$, there is one subtlety
that we do {\it not} have to face.  The $SO(32)$ element $x$ can be
lifted to $Spin(32)/\Z_2$ in two different ways (depending on the sign
of the action on spinors), but because the gauge bundle with monodromy
$x$ can be considered as an $SO(32)$ bundle, the instanton number and
eta invariant can be computed in $SO(32)$, and one does not need to
worry about the choice of lifting.  For bundles without vector
structure, the choice of lifting does matter.

Now we consider elements of order two in $Spin(32)/\Z_2$ that square to
$-1$ in $SO(32)$.  An $SO(32)$ matrix that squares to $-1$ is equivalent
to the matrix $M$.  $M$ can be lifted to $Spin(32)/\Z_2$ in two ways,
giving two group elements that we will call $M$ and $M'$.  Note that the
formula (\ref{ippo}) applies equally well to $M$ or $M'$ (as they act
the same way in the adjoint representation), so in either case the eta
invariant shifts the effective instanton number by $-1$.  However, $M$
and $M'$ have different values of the allowed instanton number.  Recall
that we constructed above an explicit instanton with monodromy $M$ and
instanton number $1$.  This was done by taking in $SO(32)$ a total of 16
commuting copies of an $SO(2)$ instanton with magnetic flux
\begin{equation}
{\int_Y{F\over 2\pi}={1\over 4}\left(\begin{array}{cc} 0 & 1 \cr -1 & 0
           \end{array} \right).}
\label{nurgo}
\end{equation}
Each $SO(2)$ factor contributed $1/16$ to the instanton number.  Now to
get monodromy $M'$ instead of $M$, we want to make at infinity an extra
$2\pi$ rotation in one of the $SO(2)$ subgroups.  This will occur if we
add one Dirac quantum to the magnetic flux (integrated over $Y$) in that
subgroup, so that one will have
\begin{equation}
{\int_Y{F\over 2\pi}={5\over 4}\left(
\begin{array}{cc}0 & 1 \cr -1 & 0 \cr\end{array}\right).}
\label{kurgo}
\end{equation}
This subgroup will now contribute $5^2/16$ to the instanton number, so in
going from $M$ to $M'$ the instanton number has changed by
$(5^2-1^2)/16=3/2$, showing that with monodromy $M'$ at infinity, the
instanton number is half-integral.

Now we move on to $E_8$.  $E_8$ actually has only two subgroups of order
two.  One of them is obtained by considering a subgroup $(SU(2)\times
E_7)/\Z_2$ of $E_8$ and taking the $\Z_2$ generated by the element $-1$
of $SU(2)$.  This breaks $E_8$ to $(SU(2)\times E_7)/\Z_2$ and gives
half-integer $k$ and $n_-=112$.  The other case is obtained by taking
the $\Z_2$ to be the center of a $Spin(16)/\Z_2$ subgroup of $E_8$,
breaking $E_8$ to $Spin(16)/\Z_2$.  This gives integer $k$ and
$n_-=128$.

To prove that these are the only $\Z_2$ subgroups of $E_8$, note that
one can assume that the generator $x$ of $\Z_2$ is in $Spin(16)/\Z_2$,
which contains a maximal torus.  If $x^2=1$ in $SO(16)$, even without
dividing by the $\Z_2$, then as above one can realize $x$ as a diagonal
$SO(16)$ matrix with $p$ eigenvalues $-1$ and $16-p$ eigenvalues $1$.
By arguments as above, the only cases that one needs to consider are
$p=4$ and $p=8$.  These can be seen to correspond respectively to the
unbroken groups $(SU(2)\times E_7)/\Z_2$ and $Spin(16)/\Z_2$.  One can
also consider the case in which in $SO(16)$, $x^2=-1$, so that $x^2=1$
only in $Spin(32)/\Z_2$.  This corresponds to the case that $x$ is a
matrix $N$ which is the sum of eight blocks of the form
\begin{equation}
{\left(\begin{array}{cc} 0 & 1 \cr -1 & 0 \end{array}\right)}
\label{kurry}
\end{equation}
(analogous to the $SO(32)$ matrix $M$ that entered earlier).  The
subgroup of $SO(16)$ left unbroken by $N$ is $U(8)$, but additional
unbroken symmetries come from generators of $E_8$ in the spinor of
$SO(16)$.  In defining the action of $N$ on the spinor representation of
$SO(16)$, there is an arbitrary minus sign, and what unbroken group one
gets depends on how this sign is chosen.  With one choice of sign one
gets the unbroken $(SU(2)\times E_7)/\Z_2$ seen earlier, while the other
choice gives another way to construct the $\Z_2$ symmetry with unbroken
$Spin(16)/\Z_2$.  We will label these two lifts of $N$ to $E_8$ as $N$
and $N'$, respectively.  This uniform construction of the two
inequivalent $\Z_2$ subgroups of $E_8$, differing only by how the action
of $N$ is lifted to spinors, will be convenient when we analyze
$T$-dualities.

\bigskip\bigskip\noindent{\it The Nonperturbative Inconsistency}

We can now obtain a better understanding of the nonperturbative
inconsistency found in section~2.3.  Thus far we have considered the
Dirac quantization condition on the small two-sphere $S$ located at each
orbifold point.  There are other closed 2-cycles on $K3$.  Consider four
fixed points lying in a plane (for example $x^6 = x^7 = 0$); label them
by $\alpha$.  There is sphere $S'$ which intersects each of the
$S_\alpha$ once.  It follows that
\begin{equation}
\int_{S'}{F\over 2\pi}= \sum_\alpha \int_{Y_\alpha}{F\over 2\pi}.
\end{equation}
For holonomy
$M$, this is just the matrix~(\ref{kurry}) in each $U(1)$ and so
is integer-valued in the positive chirality spinor.  However, if $m$ of
the fixed points have holonomy $M'$, then in one $O(2)$ one obtains
$(m+1)$ times~(\ref{kurry}), shifting the value in the spinor by $\pm
m/2$.  It follows that for $m$ odd the positive chirality spinor does
not satisfy Dirac quantization on $S'$; there is an obstruction $\tilde
w_2$.  From the discussion above, the number of instantons in the plane
is $m/2$ plus an integer, so the condition for the positive chirality
spinor to exist is that the number of instantons in each plane be an
integer.  This is slightly stronger than the ($T$-dual of the) condition
found in section~2.3.

To summarize, the nonperturbative inconsistency has a simple origin.  A
vector bundle which admits tensors but not spinors of $SO(32)$ can be a
consistent background for the Type~I string in perturbation theory, but
not nonperturbatively.

\subsection{$Spin(32)/\ZZ_2$ Instantons on {\rm K3}}

We would now like to study in more detail $Spin(32)/\Z_2$ instantons on
K3.  In addition to the instanton number $k$, a $Spin(32)/\Z_2$ bundle
$E$ on K3 is classified by the characteristic class $\tilde w_2(E)$,
which is the obstruction to $E$ admitting ``vector structure.''  $\tilde
w_2(E)$ takes values in $H^2({\rm K3},\Z_2)$, which has $2^{22}$
elements, so for a fixed K3 there are $2^{22}$ topological classes of
$E$ to consider, for given $k$.  However, if one classifies $E$'s only
up to diffeomorphism (which may be appropriate if one plans to let the
gravitational moduli of the K3 vary arbitrarily), there are only a few
cases.  In fact, being a mod two cohomology class, $\tilde w_2$ can be
lifted to an integral cohomology class that is well-defined modulo two.
Its square (which is even because K3 is a spin manifold) is therefore
well-defined modulo four.  K3 has a very large diffeomorphism group
(see for instance chapter six of~\cite{friedman}),
and it can be shown that if $\tilde w_2$ is non-zero, its only invariant
is the value of $\tilde w_2^2$ modulo four. So there are three cases:
$\tilde w_2=0$, which corresponds to the $SO(32)$ bundles that have been
assumed in the past; $\tilde w_2$ non-zero and $\tilde w_2^2$ congruent
to 2 modulo four; and $\tilde w_2$ non-zero but $\tilde w_2^2$ congruent
to 0 modulo four.  Of the three cases, we will only study two in this
paper: the conventional case with $\tilde w_2=0$, and the bundle
relevant to the GP model, which (with $\tilde w_2$ supported on
two-spheres obtained by blowing up 16 $\Z_2$ fixed points) has $\tilde
w_2\not= 0$, but $\tilde w_2^2$ congruent to 0 modulo four.

The index formula for the dimension of instanton moduli space says that
(regardless of the value of $\tilde w_2$)
\begin{equation}
{{\rm dim}\,{\cal M}_k=120k-992,}
\label{murmo}
\end{equation}
for $k$ large enough that the generic instanton completely breaks the
gauge symmetry.  Is this true for the value $k=24$ that is relevant to
K3 compactification of the $Spin(32)/\Z_2$ heterotic string?  A
necessary but not sufficient condition is that the right hand side of
(\ref{murmo}) should be positive.  For the physical value, $k=24$, the
necessary condition is obeyed, but nonetheless complete breaking of the
symmetry is not possible for the standard bundle with $\tilde w_2=0$.
This has been argued as follows.  Take an $SU(2)$ instanton of instanton
number 24, embedded in $SO(32)$ so as to break $SO(32)$ to $SU(2)\times
SO(28)$.  This gives an $SU(2)\times SO(28)$ theory with 10 copies of
the $(2,28)$ of $SU(2)\times SO(28)$, a spectrum that makes possible
Higgsing to an $SO(8)$ without charged fields.

This argument shows that there is a branch of the moduli space on which
the generic unbroken gauge group is $SO(8)$, but does not show that
there is not another branch with, for instance, complete symmetry
breaking.  The following simple argument shows this and exhibits
directly that ``vector structure'' is the key issue.

Given an $SO(32)$ bundle $E$ over K3 with instanton number $k$, let $D$
be the Dirac operator on spinors {\it with values in the vector
representation of $SO(32)$}.  One of the two spin bundles of K3 is
trivial; let us call this the bundle of positive chirality spinors.  The
index of $D$, that is the number of zero modes of positive chirality
minus the number with negative chirality, is from the index theorem
\begin{equation}
{I=2(32-k).}
\label{iggo}
\end{equation}
For $k<32$ there are thus at least $2(32-k)$ positive chirality zero
modes.  Let $\psi$ be such a mode.  Using the fact that the curvature of
K3 and of the instanton bundle are both anti-self-dual
and that $\psi $ has
positive chirality, one gets
\begin{equation}
{0=\int_{{\rm K3}} (D\psi,D\psi)=\int_{{\rm K3}}(\psi,D^2\psi)
=-\int_{{\rm K3}}(\psi,D_\alpha
D^\alpha \psi)=\int_{{\rm K3}}(D_\alpha\psi,
D^\alpha\psi),}
\label{liggo}
\end{equation}
with the gamma matrix terms canceling out (since for instance $\bar\psi
\Gamma^{\alpha\beta}F_{\alpha\beta}\psi=0$ because of considerations of
self-duality and chirality).  So such a zero mode is covariantly
constant, $D_\alpha\psi=0$.  Since the positive chirality spin bundle of
K3 is of rank two, to get $2(32-k)$ covariantly constant positive
chirality spinors, the structure group of the gauge connection on $E$
must leave fixed a $32-k$ dimensional subspace of the vector
representation of $SO(32)$.  So for any $k\leq 30$, there is always an
unbroken $SO(32-k)$.

This argument clearly uses heavily the existence of vector structure,
and one may ask what happens for $\tilde w_2\not=0$.  For the $k=24$
bundle relevant to the GP model, complete Higgsing is possible; this is
clear from the explicit spectrum found in ref.~\cite{gimpol}.  This is
also true for a $k=24$ bundle with $\tilde w_2^2$ congruent to two
modulo four.  In fact, one can construct such a bundle with 23 $SO(32)$
instantons, breaking to $SO(9)$; the last instanton can be a
$Spin(32)/\Z_2$ instanton, supported on a two-sphere $S$ of $S\cdot
S=-2$, as constructed above, and breaking $Spin(32)/\Z_2$ to $U(16)$.
If the $SO(9)$ and $U(16)$ are aligned generically in $Spin(32)/\Z_2$,
their intersection is trivial, showing that complete Higgsing is
possible.

\subsection{Comparison to $E_8$ Instantons}

Upon toroidal compactification, the $SO(32)$ heterotic string is
equivalent to the $E_8\times E_8$ heterotic string.  One may ask
to what extent that is also true upon K3 compactification.

In $E_8\times E_8$ compactification on K3, one may place $12+n$
instantons in one $E_8$ and $12-n$ in the other, with $0\leq n \leq 12$.
The generic unbroken gauge symmetry depends on $n$.  One generically has
complete Higgsing for $n=0,1,2$, while for $n>2$ there is a generic
unbroken gauge group.  The $n=0 $ and $n=2$ models are actually
equivalent.  This is known from $F$-theory~\cite{vafamorrison}, though
it is conceivable that there might exist a more down-to-earth
explanation via some sort of $T$-duality.

We would like to know whether the various $Spin(32)/\Z_2$ models are
equivalent to some of the $E_8\times E_8$ models.  The conventional
model based on the $SO(32)$ bundle with vector structure has -- as we
have explained above -- a generic unbroken $SO(8)$ symmetry, with no
massless charged hypermultiplets.  The $E_8\times E_8$ model with those
characteristics is the $n=4$ model, so one is led to conjecture that (as
independently suggested in \cite{vafamorrison}) the $Spin(32)/\Z_2$
model with vector structure is equivalent to the $E_8\times E_8$ model
with instanton numbers $(16,8)$.  Later we will demonstrate, via
$T$-duality, that this is so.

We would also like to identify GP models with $E_8\times E_8$ models.
In particular, a special case of the GP models, as explained in section
4.2, has a description as a $Spin(32)/\Z_2$ heterotic string orbifold.
This model has the gauge group completely broken generically, so the
$E_8\times E_8$ models to which it might be equivalent are $n=0$ and 1.
In fact, we will argue later using $T$-duality that this particular
$Spin(32)/\Z_2$ model is equivalent to the $n=0$ model, that is to the
$E_8\times E_8$ model with equal instanton numbers in the two $E_8$'s.

\bigskip\noindent{\it Comparison to $E_8\times E_8$ Perturbation Theory}

This last-mentioned equivalence actually makes it possible to resolve
some puzzles that were left hanging in \cite{dmw}.  We will pause to
explain this here, before going on in the next subsection to construct
the $T$-dualities by which the $Spin(32)/\Z_2$ and $E_8\times E_8$
models can be related.  In \cite{dmw}, it was shown that the $E_8\times
E_8$ model with $n=0$, that is with instanton numbers $(12,12)$, has a
strong-weak coupling duality that exchanges perturbative massless gauge
fields, which arise via conventional symmetry restoration when Higgs
expectation values are turned off, with massless gauge fields that arise
non-perturbatively at singularities.

An attempt was made in section 4 of \cite{dmw} to match particular
unbroken gauge groups with particular singularities, associated with
small instantons.  The paradox is that extensive numerical evidence was
offered for such matching, which however was based on assumptions that
did not appear sound.

For example, it was proposed that un-Higgsing of a perturbative $Sp(n)$
gauge group for $n=1,2,3$ was dual to the collapse of $n$ small
instantons at a point.  This proposal neatly fits the facts {\it if} the
small instantons in question generate just the non-perturbative gauge
groups that actually arise \cite{sminst} from small $Spin(32)/\Z_2$
instantons.  Since the model in question was an $E_8\times E_8$ model,
it was assumed in \cite{dmw} that the small instantons would have to be
small $E_8$ instantons.  But it is now clear
\cite{hanany,seiwit} that small $E_8$ instantons behave in a
very different (and more exotic) way than small $Spin(32)/\Z_2 $
instantons.  The fact that the $(12,12)$ $E_8\times E_8$ model turns out
to be equivalent to a $Spin(32)/\Z_2$ model resolves the contradiction.
Because of this relation, in addition to singularities due to small
$E_8$ instantons, the model also has singularities due to small
$Spin(32)/\Z_2$ instantons.  The facts presented in \cite{dmw} all fit
neatly if we reinterpret the claim to be that the restoration of a
perturbative $Sp(n)$ symmetry is dual to the collapse of $n$ coincident
$Spin(32)/\Z_2$ instantons.

Another contradiction in \cite{dmw} concerned the special case of $n=3$,
that is un-Higgsing of an $Sp(3)$ subgroup.  This should be dual to
collapse of three instantons at a point.  If they are $E_8$ instantons,
then in the $(12,12)$ model, this leaves only 9 instantons in one of the
two $E_8$'s, which would lead to the un-Higgsing of a perturbative
$SU(3)$ group that is not predicted by the duality.  Resolutions of this
puzzle suggested in \cite{dmw} were not very convincing, but the
situation is now clear: the instantons in question are $Spin(32)/\Z_2 $
instantons, and collapse of three of them need not lead to restoration
of any perturbative gauge symmetry.

We can likewise now resolve a number of contradictions in the discussion
in \cite{dmw} of the restoration of an $SU(n)$ gauge symmetry, for
$n=3,\dots,6$.  This was interpreted in terms of the collapse of $n/2$
instantons at a $\Z_2$ orbifold singularity.  The difficulty here is
that in the numerical evidence in \cite{dmw}, it was necessary to claim
that the moduli space of $E_8$ instantons of instanton number $n/2$ on
the Eguchi-Hansen space has dimension $60n$, for even or odd $n$.  While
this is true for even $n$ (provided we take the monodromy at infinity to
be trivial -- recall that there is another choice that gives integral
instanton number), it is, because of the correction involving the eta
invariant, false for odd $n$.  The correction is $-n_-/2=-56$, given
that $n_-=112$ for the monodromy at infinity that gives half-integral
instanton number.

If, however, one reinterprets the discussion in terms of $Spin(32)/\Z_2$
instantons, then all becomes clear.  We consider in the neighborhood of
a $\Z_2$ orbifold singularity a bundle without vector structure (since
that is the sort of bundle that arises in the GP model near orbifold
singularities, as we have discussed).  The monodromy at infinity can
therefore be $M$ or $M'$, corresponding as we have seen to integer or
half-integer instanton number.  The eta invariant is not zero, but has
the effect, as we calculated in (\ref{ippo}), of shifting the instanton
number by one.  Thus, we modify the proposal in \cite{dmw} to assert
that the un-Higgsing of an $SU(n)$ perturbative gauge symmetry, for
$n=3,\dots,6$, is dual to collapse of $1+n/2 $ $Spin(32)/\Z_2$
instantons at a $\Z_2$ orbifold singularity without vector structure.
Not only is this revised proposal free of the previous contradictions,
but (given the $T$-duality that we will discuss in the next subsection)
it is indeed in agreement with the spectrum found in ref.~\cite{gimpol}.
There it was found (without restriction to $n\leq 6$) that with $n/2$
five-branes (equivalent to $2n$ Chan-Paton indices) at an $A_1$ orbifold
singularity without vector structure, one gets an $SU(n)$ gauge
symmetry.  We claim that there is an instanton hidden in this kind of
orbifold singularity even before five-branes come near, so the total
instanton number carried by the orbifold singularity to give $SU(n)$
symmetry is indeed $1+n/2$.

\subsection{$T$-Dualities between K3 Compactifications}

Here we will, finally, justify the claims made earlier about
$T$-dualities between certain K3 compactifications of
the $Spin(32)/\Z_2$ and $E_8\times E_8$ heterotic strings.

It is helpful first to recall how the $Spin(32)/\Z_2$ and $E_8\times
E_8$ theories are related after compactification on $\R^9\times \S^1$.
One can interpolate in three steps from the vacuum with unbroken
$Spin(32)/\Z_2$ to the vacuum with unbroken $E_8\times E_8$:

{\it (a)}  Starting
with unbroken $Spin(32)/\Z_2$ with a very large $\S^1$,
one continuously
turns on a Wilson line that breaks this group to a subgroup
that is locally $SO(16)\times SO(16)$.  The requisite Wilson line
is a diagonal matrix $W={\rm diag}(1,\dots, 1, -1,\dots, -1)$ with
16 eigenvalues 1 and 16 eigenvalues $-1$.

{\it (b)} Then one makes an $r\to 1/r$ transformation on the $\S^1$.
In other words, one continuously reduces the radius of the $\S^1$ until
it is very small; at that point, the theory is better described via
a $T$-duality transformation which makes $r$ large again.
The dual theory is an $E_8\times E_8$ theory with a Wilson line $W'$
that breaks $E_8\times E_8$ to a subgroup that is locally $SO(16)\times
SO(16)$.  The Wilson line in question is a product, in each $E_8$,
of the group element $(-1)^F$ corresponding to a $2\pi$ rotation
in an $SO(16)$ subgroup; this group element acts as $+1$ in the part of
the adjoint representation of $E_8$ that transforms as the adjoint of
$SO(16)$, and $-1$ on the rest.

{\it (c)}  Finally, one can continuously turn off the Wilson line $W'$
and restore the $E_8\times E_8$ gauge symmetry.

Note that, though all three steps are needed to interpolate from
unbroken $Spin(32)/\Z_2$ to unbroken $E_8\times E_8$, step {\it (b)} is
the only step that is really necessary to show that the $Spin(32)/\Z_2$
and $E_8\times E_8$ theories are equivalent.  Step {\it (b)} shows by
itself that the $Spin(32)/\Z_2$ theory with the gauge group weakly
broken by a certain Wilson line (Wilson line symmetry breaking is weak
if $r$ is large) is continuously connected
to the $E_8\times E_8$ theory with the
gauge group similarly weakly broken.  We mention this because, when we
get to K3 orbifolds, there may be an obstruction to steps {\it (a)} or
{\it (c)}; it is only step {\it (b)} that we need to implement.

\bigskip\noindent{\it Model with Vector Structure}

In actually studying $T$-dualities between $Spin(32)/\Z_2$ and
$E_8\times E_8$ heterotic strings, we will consider K3  manifolds
constructed as $\Z_2$ orbifolds of a four-torus $\T^4$.  On the
four-torus we will sometimes have Wilson lines.

When one divides the four-torus by $\Z_2$, one must also make a $\Z_2$
twist in the gauge group in order to preserve level matching.  We first
consider the case of a $Spin(32)/\Z_2$ model with vector structure, so
that the generator $x$ of the $\Z_2$ twist can be regarded as an element
of $SO(32)$ (but $x$ is equivalent to $-x$ because the gauge group is
really $Spin(32)/\Z_2$).  In the fermionic construction of the
$Spin(32)/\Z_2$ heterotic string, the $SO(32)$ is carried by 32
left-moving Majorana-Weyl fermions, and the twist simply multiplies $p$
of them by $-1$; because $x$ is equivalent to $-x$ we can take $p\leq
16$.  For level matching, $p$ must be congruent to 4 modulo 8, so the
possibilities are $p=4$ and $p=12$.  $p=4$ corresponds to the ``standard
embedding of the spin connection in the gauge group'' (and breaks
$SO(32)$ to $ SO(28)\times SO(4)$, which becomes $SO(28)\times SU(2)$ if
one replaces the orbifold by a smooth K3), but we will here consider the
other case $p=12$.

The $p=12$ model has unbroken $SO(20)\times SO(12)$ (where the two
factors act respectively on left-moving fermions that are even or odd
under $x$).  The spectrum of massless hypermultiplets consists of a
$({\bf 20,12})$ from the untwisted sector and sixteen $({\bf 1,32})$
half-hypermultiplets from twisted sectors (the {\bf 32} is a chiral spinor of
$SO(12)$).  The unbroken gauge group of this model after generic
Higgsing (at least on one obvious branch) is easily determined.  After
using the $({\bf 1,32})$'s to
completely break the $SO(12)$, the $({\bf 20,12})$
suffices to break $SO(20)$ down to $SO(8)$.  So the model, at least on
this branch, has a generic unbroken $SO(8)$, as expected for a K3
compactification of the $Spin(32)/\Z_2$ heterotic string with vector
structure.

Now we want to turn on a Wilson line $W$ that breaks $SO(32)$ to
$SO(16)\times SO(16)$ (and together with $x$ breaks $SO(32)$ to a
smaller group).  To continuously turn on $W$, we write
\begin{equation}
{W=\exp(\pi b),}
\label{ewer}
\end{equation}
where $b$ is an $SO(32)$ generator that is conjugate to
eight copies of
\begin{equation}
{\left(\begin{array}{cc} 0 & 1 \cr -1 & 0 \end{array}\right)}
\label{funny}
\end{equation}
plus a $16\times 16$ identity matrix.  Since $x$ multiplies all four
coordinates of the four-torus by $-1$, we can continuously
turn on $W$ by setting $W_t=\exp(\pi tb)$, $0\leq t\leq 1$, provided
\begin{equation}
{ xb=-bx .}
\label{punny}
\end{equation}
This can be achieved as follows.  Break up the 32 fermions of the
$Spin(32)/\Z_2$ heterotic string into two groups of $16$, where $x$ has
four eigenvalues $-1$ in the first group and eight in the second group.
Take $b$ to be zero in the first group, and in the second group (in a
basis in which $x$ is $-1$ on the first eight basis elements and $+1$ on
the others) take $b$ to be the matrix
\begin{equation}
b=\left(\begin{array}{cc}0 & I \cr -I & 0 \cr\end{array}\right)
\label{kkk}
\end{equation}
in eight by eight blocks.  This $b$ has the desired properties.

So starting with this $\Z_2$ orbifold of the $Spin(32)/\Z_2$ heterotic
string, we can continuously turn on the Wilson line $W$, implementing
step {\it (a)} in the above scenario.  Then we can implement the crucial
step {\it (b)}, making an $r\to 1/r$ transformation on the circle that
has the Wilson line, thereby mapping the $Spin(32)/\Z_2$ model to an
$E_8\times E_8$ model.

The $E_8\times E_8$ heterotic string has a convenient fermionic
construction with two groups of 16 left-moving free fermions and a
separate GSO-like projection on each.  When the $E_8\times E_8$ model is
obtained as just described, the matrix $x$ acts with four eigenvalues
$-1$ on the first group and with eight on the second.  Such elements of
$E_8$ were discussed in our classification of $\Z_2$ subgroups of $E_8$,
and break $E_8$ to $SU(2)\times E_7$ and $SO(16)$, respectively.  After
arriving at an $E_8\times E_8$ model in this fashion, we also have a
Wilson line that would break $E_8\times E_8$ to $SO(16)\times SO(16)$
and breaks $SU(2)\times E_7\times SO(16)$ to $SU(2)\times SU(2)\times
SO(12)\times SO(8)\times SO(8)$.  (Note that the unbroken subgroup of
the second $E_8$ is $SO(8)\times SO(8)$, since this is the intersection
of the two $SO(16)$'s that commute respectively with $x$ and with $W$.)

One may ask whether the Wilson line symmetry breaking can be turned off
to get a pure $\Z_2$ orbifold of the $E_8\times E_8$ heterotic string.
To do this, one wants to write in each $E_8$
\begin{equation}
{(-1)^F=e^{\pi b},}
\label{jumbo}
\end{equation}
where $b$ is an $E_8$ generator with $xb=-bx$; this enables
one to interpolate from the 1 to $(-1)^F$ via $W_t=\exp(\pi tb)$, as
before.  In a basis in which the
$-1$ eigenvalues of $x$ are the first four of the first group of sixteen
fermions and the first eight in the second group, one can take $b$ to
generate a rotation of the 4-5 plane of the first group and the 8-9
plane of the second.

\bigskip\noindent{\it Model without Vector Structure}

Now we want to relate $E_8\times E_8$ compactification with
instanton numbers $(12,12)$ to a $Spin(32)/\Z_2$ model without
vector structure.

It will be convenient to start on the $E_8\times E_8$ side.
The way we will ensure that an orbifold corresponds to a $(12,12)$
compactification is by showing that it is symmetric between the two
$E_8$'s.

There is actually no $\Z_2$ orbifold of the four-torus that is symmetric
in the two $E_8$'s.  To see this, use the fermionic construction of
$E_8\times E_8$, with two groups of sixteen fermions. The two possible
$\Z_2$ generators of $E_8$, say $N$ and $N'$, can be realized by
matrices that act as $-1$ on four or eight fermions, respectively,
breaking $E_8$ to $SU(2)\times E_7$ or $SO(16)$.  Level matching
requires that the number of twisted fermions is 4 modulo 8, so the
possible $\Z_2$ twists are $N\times 1$ (the standard embedding of the
spin connection in the gauge group, with instanton numbers $(24,0)$) or
$N\times N'$, with unbroken gauge group $SU(2)\times E_7\times SO(16)$.
No level-matched choice is symmetric between the two $E_8$'s.

There is, however, a $\Z_2$ orbifold with an additional $\Z_2$ Wilson
line (one might describe it as a $\Z_2\times \Z_2$ orbifold) that is
symmetric in the two $E_8$'s.  (This was pointed out independently in
\cite{ibanez}.)  To see this, use another construction of $N$ and $N'$
that was described in our classification of $\Z_2$ subgroups.  In this
construction, $N$ is an $SO(16)$ matrix that breaks $SO(16)$ to $U(8)$,
and $N'=(-1)^FN$.

Consider a $\Z_2$ orbifold of $\T^4$ with the $\Z_2$ acting on gauge
fermions by $N\times N'$.  This is equivalent to the model described two
paragraphs ago with unbroken gauge group $SU(2)\times E_7\times SO(16)$,
though the unbroken subgroup of $SO(16)\times SO(16)$ is only
$U(8)\times U(8)$.  It is not symmetric in the two $E_8$'s.  Consider,
however, a related model with in addition a $\Z_2 $ Wilson line
$(-1)^F\times (-1)^F$ in, say, the $x^{10}$ direction, which we take to
have period one.

Thus, the transformation $x^{10}\to -x^{10}$ (with also inversion of
$x^7,\dots, x^9$) acts in $E_8\times E_8$ as $N\times N'$.  The
transformation $x^{10}\to x^{10}+1$ acts as $(-1)^F\times (-1)^F$.  The
combined transformation $x^{10}\to -x^{10}+1$ therefore acts as
$N(-1)^F\times N'(-1)^F$.  But this equals $N'\times N$, which differs
{}from $N\times N'$ by exchange of the two $E_8$'s.  Moreover,
$x^{10}\to -x^{10}+1$ is conjugated to $x^{10}\to -x^{10}$ by the
symmetry $x^{10}\to x^{10}+1/2$.  The conclusion, then, is that this
particular $\Z_2\times \Z_2$ orbifold (or $\Z_2$ orbifold with Wilson
line) is invariant under exchange of the two $E_8$'s accompanied by
$x^{10}\to x^{10}+1/2$.  Therefore, it corresponds to a K3
compactification with equal instanton numbers $(12,12)$.  (One can
verify that the massless hypermultiplet spectrum makes complete Higgsing
possible.)

Moreover, because the Wilson line we used is just the one that by itself
would break $E_8\times E_8$ to $SO(16)\times SO(16)$, we are in the
right situation to interpolate to $Spin(32)/\Z_2$: all we need to do is
to make the usual $r\to 1/r$ transformation in the direction with the
Wilson line.  In the resulting $Spin(32)/\Z_2$ model, the twist $N\times
N'$ (which is represented on the 32 fermions by the direct sum of 16
copies of the $2\times 2$ matrix in (\ref{funny})) is our friend $M$, the
$\Z_2$ twist of $Spin(32)/\Z_2$ that forbids vector structure.  So we
have succeeded in showing that the $E_8\times E_8$ compactification with
equal instanton numbers is equivalent to a $Spin(32)/\Z_2$
compactification without vector structure.

\sect{Acknowledgments}

The research of MB, RGL and NS was supported in part by DOE grant
\#DE-FG02-96ER40559; that of JP by NSF grants \#PHY91-16964 and
\#PHY94-07194; that of JHS by DOE grant
\#DE-FG03-92-ER40701; and that of EW by NSF grant \#PHY95-13835.

\pagebreak


\begin{thebibliography}{99}

\bibitem{sminst}E. Witten, ``{\it Small Instantons in String Theory,}''
\npb{460}{1996}{541}, hep-th/9511030.

\bibitem{horava}P. Horava and E. Witten,  ``{\it Heterotic and Type I
String Dynamics from Eleven Dimensions,}''  \npb{460}{1996}{506},
hep-th/9510209; ``{Eleven-Dimensional Supergravity on a Manifold with
Boundary,}'' IASSNS-HEP-96-17, hep-th/9603142.

\bibitem{witusc}
E. Witten, ``{\it String Theory Dynamics In Various Dimensions,}''
\npb{433}{1995}{85}, hep-th/9503124.

\bibitem{vafaf}C. Vafa, ``{\it Evidence for F Theory,}''
HUTP-96-A004, hep-th/9602022.

\bibitem{seiwit}N. Seiberg and E. Witten, ``{\it Comments on String
Dynamics in Six Dimensions,}'' RU-96-12, hep-th/9603003.

\bibitem{dlp}M. Duff, H. Lu and C.N. Pope, ``{\it Heterotic Phase
Transitions and Singularities of The Gauge Dyonic String,''}
CTP-TAMU-9-96, hep-th/9603037.

\bibitem{gimpol}E.G. Gimon and J. Polchinski,
``{\it Consistency Conditions for Orientifolds and D-Manifolds,}''
NSF-ITP-96-01, hep-th/9601038.

\bibitem{dmw}M.J. Duff, R. Minasian and E. Witten,
``{\it Evidence for Heterotic/Heterotic Duality,}''
CTP-TAMU-54/95, hep-th/9601036.


\bibitem{gsmech}M.B. Green and J.H. Schwarz,
``{\it Anomaly Cancellations in Supersymmetric $D=10$ Gauge Theory and
Superstring Theory,}'' \plb{149}{1984}{117};
``{\it Infinity Cancellation in $SO(32)$ Superstring Theory,}''
\plb{151}{1985}{21};
``{\it The Hexagon Gauge Anomaly in Type I Superstring Theory,}''
\npb{255}{1985}{93}.

\bibitem{dsw} E. Witten, ``{\it Some Properties of $O(32)$
Superstrings,}''
\plb{149}{1984}{351}; M. Dine, N. Seiberg, and E. Witten, ``{\it
Fayet-Iliopoulos Terms in String Theory,}'' \npb{289}{1987}{589}.

\bibitem{pcj}
J. Polchinski, S. Chaudhuri, and C. V. Johnson, ``{\it Notes on
D-Branes,}'' NSF-ITP-96-03, hep-th/9602052.

\bibitem{duff}
M. J. Duff, ``{\it Strong/Weak Coupling Duality from the Dual String,}''
\npb{442}{1995}{47}, hep-th/9501030.

\bibitem{dbrane}J. Polchinski,
``{\it Dirichlet-Branes and Ramond-Ramond Charges,}''
\prl{75}{1995}{4724}, hep-th/9510017;
J. Dai, R.G. Leigh and J. Polchinski,
``{\it New Connections between String Theories,}'' \mpl{4}{1989}{2073}.

\bibitem{vafamorrison}
D. Morrison and C. Vafa, ``{\it Compactifications of F-Theory on
Calabi-Yau Three-Folds -- I and II,}''
DUKE-TH-96-106, hep-th/9602114, and DUKE-TH-96-107, hep-th/9603161.

\bibitem{alvwit}L. Alvarez-Gaum\' e and E. Witten,
``{\it Gravitational Anomalies,}''
\npb{234}{1983}{269}.

\bibitem{gsw}M.B. Green, J.H. Schwarz and E. Witten, ``{\it Superstring
Theory,}''  Cambridge University Press, 1987.

\bibitem{bound}
M. Li, ``{\it Boundary States of D-Branes and Dy-Branes,}''
\npb{460}{1996}{351}, hep-th/9510161;
M. Douglas, ``{\it Branes Within Branes,}'' RU-95-92, hep-th/9512077.

\bibitem{adi}
C. Vafa and E. Witten, ``{\it Dual String Pairs with $N=1$ and $N=2$
Supersymmetry in Four Dimensions,}''
HUTP-95-A023, hep-th/9507050.

\bibitem{sag}
A. Sagnotti, ``{\it A Note on the Green-Schwarz Mechanism in Open-String
Theory,}'' \plb{294}{1992}{196}.

\bibitem{polwit}
J. Polchinski and E. Witten, ``{\it Evidence for Heterotic -- Type I
Duality,}'' \npb{460}{1996}{525}, hep-th/9510169.


\bibitem{douglasmoore}
M.R. Douglas and G. Moore,
``{\it D-branes, Quivers and ALE Instantons}'' hep-th/9603167.

\bibitem{friedman}
R. Friedman and J. W.
Morgan, ``{\it Smooth Four-Manifolds And Complex Surfaces,}''
(Springer-Verlag, 1994).)

\bibitem{hanany}
O. J. Ganor and A. Hanany, ``{\it Small $E_8$ Instantons and Tensionless
Noncritical Strings,}'' IASSNS-HEP-96-12, hep-th/9602120.

\bibitem{ibanez}
G. Aldazabal, A. Font, L.E. Iba\~nez, and F. Quevedo,
``{\it Heterotic/Heterotic Duality in $D=6$ and $D=4$,}''
CERN-TH-96-35, hep-th/9602097.

\end{thebibliography}
\end{document}